\documentclass[iicol]{sn-jnl}

\usepackage[utf8]{inputenc}
\usepackage[T1]{fontenc}
\usepackage{lmodern}
\usepackage{diagbox}
\usepackage{ltl}
\usepackage{amssymb}
\usepackage{amsmath}
\usepackage{stmaryrd}
\usepackage{tikz}
\usepackage{xcolor}
\usetikzlibrary{backgrounds, positioning, arrows, calc}
\usepackage{mathtools}
\usepackage[inline]{enumitem}  
\usepackage{cleveref}
\usepackage{lineno}
\usepackage{multirow}
\usepackage{wrapfig}
\usepackage[font=small]{caption}
\usepackage{subcaption}
\usepackage{hyperref}
\usepackage{graphicx}
\usepackage{amsfonts}
\usepackage{mathrsfs}

\hypersetup{
	colorlinks=true,
	allcolors=blue!70!black,
	pdfstartview=Fit,
	breaklinks=true,
}

\definecolor{admAccentBlue}{RGB}{73,111,136}
\definecolor{admAccentOchre}{RGB}{154,112,64}
\definecolor{admAccentRed}{RGB}{146,86,82}
\definecolor{admNeutralLine}{RGB}{74,74,74}

\renewcommand{\cref}{\Cref}
\renewcommand{\pfx}{\textsf{prefix}}
\newcommand{\first}{\textsf{first}}
\newcommand{\last}{\textsf{last}}
\renewcommand{\sfx}{\textsf{suffix}}
\newcommand{\infx}{\textsf{infix}}
\newcommand{\hb}{\rightsquigarrow}

\newcommand{\fr}{\mathsf{fr}}
\newcommand{\tr}{\mathsf{Tr}}
\newcommand{\AP}{\mathsf{AP}}
\newcommand{\true}{\texttt{true}}
\newcommand{\false}{\texttt{false}}
\newcommand{\ccf}{\mathsf{ccf}}
\newcommand{\destutter}{\mathsf{destutter}}
\newcommand{\stutter}{\mathsf{stutter}}
\newcommand*\BitAnd{\mathbin{\&}}
\newcommand*\BitOr{\mathbin{|}}
\newcommand*\BitNeg{\ensuremath{\mathord{\sim}}}
\newcommand{\R}{\mathbb{R}}
\newcommand{\B}{\mathbb{B}}
\renewcommand{\LTLf}{\LTLdiamond}
\let\LTLfinally\LTLf
\let\LTLeventually\LTLf
\renewcommand{\LTLg}{\LTLsquare}

\renewcommand{\LTLo}{\LTLcircle}

\def\until{\kern.1em\mathcal{U}}
\def\since{\kern.1em\mathcal{S}}
\newcommand{\suchthat}{\;\ifnum\currentgrouptype=16 \middle\fi|\;}
\let\st\suchthat
\newcommand{\defeq}{\coloneqq}
\newcommand{\Alt}{\mathit{Alt}}

\newcounter{resq}[section]
\newenvironment{resq}[1][]{%
	\refstepcounter{resq}%
	\par\medskip
	\noindent \textbf{RQ\theresq.}~\textit{#1} \rmfamily}{\medskip}

\newcommand{\prevquestion}{RQ\the\numexpr\value{resq}-1\relax\xspace}

\tikzstyle{state}=[thick,minimum size=18pt, circle,draw]
\tikzstyle{transition}=[->,thick,>=stealth,shorten >=1pt,shorten <=1pt]
\tikzstyle{final}=[after node path={ node[state, scale=.8] at (\tikzlastnode) {} }]
\tikzstyle{initial}=[after node path={
	[to path={[transition] (\tikztostart) -- (\tikztotarget)}]
	(\tikzlastnode)++(180:22pt) edge (\tikzlastnode)
}]
\tikzset{
	bg/.default={},
	bg/.style={execute at end picture={
			\begin{scope}[on background layer]
				\node[xshift=-1mm, yshift=-1mm] (sw) at (current bounding box.south west) {};
				\node[xshift=1mm, yshift=1mm] (ne) at (current bounding box.north east) {};
				\node[xshift=1mm, yshift=-1mm] (nw) at (current bounding box.north west) {};
				\fill[fill=black!10,rounded corners] (sw) rectangle (ne);
				
				\ifx&#1&\else
				\node[anchor=north east, xshift=2pt] at (nw) {#1};
				\fi
			\end{scope}
	}},
}

\newtheorem{theorem}{Theorem}
\newtheorem{lemma}{Lemma}
\newtheorem{corollary}{Corollary}
\newtheorem{example}{Example}

\newtheorem{definition}{Definition}

\begin{document}
	\title[Approximate Distributed Monitoring under Partial Synchrony]{Approximate Distributed Monitoring under Partial Synchrony: Balancing Speed \& Accuracy}
	\author[1]{\fnm{Borzoo} \sur{Bonakdarpour}}\email{borzoo@msu.edu}
	\author[1]{\fnm{Anik} \sur{Momtaz}}\email{momtazan@msu.edu}
	\author[2]{\fnm{Dejan} \sur{Ni\v{c}kovi\'{c}}}\email{dejan.nickovic@ait.ac.at}
	\author*[3]{\fnm{N. Ege} \sur{Sara\c{c}}}\email{ege.sarac@cispa.de}
	\affil[1]{\orgname{Michigan State University}}
	\affil[2]{\orgname{AIT Austrian Institute of Technology}}
	\affil*[3]{\orgname{CISPA Helmholtz Center for Information Security}}
		
	\abstract{
		In distributed systems with processes that do not share a global clock, \emph{partial synchrony} can be achieved by clock synchronization that guarantees bounded clock skew among all processes.
		Existing solutions for distributed runtime verification under partial synchrony against temporal logic specifications are exact but suffer from significant 
		computational overhead.
		In this paper, we propose an \emph{approximate} distributed monitoring algorithm for Signal Temporal Logic (STL) that mitigates this issue by abstracting away potential interleaving behaviors.
		This conservative abstraction enables a significant speedup, albeit with a tradeoff in precision.
		We address this tradeoff by combining our approximate monitor with its exact counterpart, gaining efficiency without sacrificing precision.
		Our framework supports \emph{offline} monitoring of both future and past STL, as well as \emph{online} monitoring of past STL.
		We evaluate our approach with multiple experiments, showcasing its efficacy in 
		both real-world applications and synthetic examples.
	}
	
	
	\maketitle

	\section{Introduction} 
	\label{sec:introduction}
	
	\emph{Distributed systems} are networks of independent agents cooperating toward a common objective.
	They take many forms, e.g., cloud computing provides computing and storage over the internet, and drone swarms coordinate tasks such as search and rescue or delivery.
	Agents in a distributed system usually do not share a global clock; their local clocks must therefore be synchronized.
	Perfect synchronization is not possible in practice because messages take time and processes may fail, but protocols such as the Network Time Protocol (NTP) can bound the difference between local clocks.
	We refer to this bounded-skew setting as \emph{partial synchrony}.

	Formal verification of distributed systems is notoriously hard because of the combinatorial explosion of interleavings among the behaviors of individual agents.
	\emph{Runtime verification (RV)} offers a more pragmatic approach in which a monitor observes an execution of a distributed system and checks it against a formal specification.
	In \emph{offline} monitoring, the monitor evaluates a complete distributed trace after execution, which is natural for analyzing log files.
	In \emph{online} monitoring, the monitor instead receives the trace incrementally in portions and issues a verdict after evaluating each new portion, typically under strict resource constraints.

	We consider the \emph{distributed RV} setting in which this task is performed by a single central monitor observing the independent agents (as opposed to \emph{decentralized RV}, where the monitoring task itself is distributed).
	Related work includes distributed RV in the fully {\em synchronous}~\cite{ef20,cf16,bf16} and {\em asynchronous}~\cite{cgnm13,mg05,og07,mb15,g20,bfrrt22} settings, as well as benchmarking tools~\cite{aafi21} for monitoring overhead.
	Under partial synchrony, distributed RV has been studied for Linear Temporal Logic (LTL)~\cite{GangulyMB20} and Signal Temporal Logic (STL)~\cite{MomtazAB23}.
	In this setting, bounded clock skew introduces uncertainty about the relative ordering of events generated by different agents, in both offline and online monitoring.
	Existing exact solutions use Satisfiability-Modulo-Theory (SMT) solving to provide sound and complete monitoring procedures.
	However, even though RV analyzes only a single execution, a distributed behavior may still admit an excessive number of possible interleavings.
	As a result, exact distributed monitors can incur significant computational overhead.
	This has been observed even under partial synchrony~\cite{GangulyMB20,GangulyMB24}, and becomes problematic even for offline monitoring when the log files are large.

	To mitigate this issue, we propose an \emph{approximate} approach to distributed RV of STL under partial synchrony: offline for both future- and past-time STL, and online for past-time STL.
	In essence, we conservatively abstract away potential interleavings in distributed behaviors, resulting in an overapproximation of them.
	This abstraction represents distributed behaviors by sets of \emph{value expressions} that account for regions of uncertainty created by clock skew.
	We define monitoring operations that evaluate temporal specifications over these expressions, yielding monitoring verdicts on overapproximated behaviors.
	Because the abstraction is conservative, conclusive verdicts remain sound, but the approach introduces a tradeoff between {\em precision} and {\em speedup}.
	For applications where reduced precision is unacceptable, we devise a methodology that combines approximate and exact monitors to benefit from the enhanced efficiency without sacrificing precision.
	Approximate monitoring is also valuable in the sequential setting, with applications including monitoring with state estimation \cite{StollerBSGHSZ11,BartocciG13}, quantitative monitoring and its resource-precision tradeoffs \cite{HenzingerS21,HenzingerMS22,HenzingerMS23}, and various other uses \cite{AlechinaDL14,AcetoAFIL21}.

	We implemented our approach in a prototype tool and performed a thorough evaluation on synthetic benchmarks and real-world case studies (mutual separation in a swarm of drones and a water distribution system).
	In many experiments, our approximate monitors achieve speedups of up to six orders of magnitude compared to the exact SMT-based solution.
	We empirically characterize the classes of specifications and behaviors for which our approximate monitors achieve good precision.
	We finally show that combining exact and approximate distributed RV yields significant efficiency gains on average without sacrificing precision, even when the approximate monitors alone do not have high precision.

	This journal version extends our earlier conference paper \cite{BonakdarpourMNS24}.
	First, in the offline setting, we improve our algorithm for evaluating timed formulas and provide the correctness proofs and additional experiments omitted from the conference version due to space constraints.
	Second, in the online setting, we extend our framework to support past-time STL and introduce two monitoring algorithms, together with their correctness proofs and a comprehensive experimental evaluation.

	\paragraph*{Organization.}
	Section~\ref{sec:preliminaries} presents the necessary background.
	Section~\ref{sec:semantics} gives an overview of our two-step monitoring approach.
	Sections~\ref{sec:approach} and~\ref{sec:algorithm} then present the trace overapproximation and the monitoring algorithms, respectively.
	We evaluate our approach in Section~\ref{sec:experiments}.
	Finally, we conclude in Section~\ref{sec:conclusion}.

	\section{Preliminaries}
	\label{sec:preliminaries}
	
	We denote by $\B \defeq \{0, 1\}$ the set of Booleans, $\R$ the set of reals, $\R_{\geq 0}$ the set of nonnegative reals, and $\R_{> 0}$ the set of positive reals.
	An interval $I \subseteq \R$ with endpoints $a < b$ has length $b - a$.
	Unless stated otherwise, all intervals in the paper are assumed to be nonempty.
	
	Let $\Sigma$ be a finite \emph{alphabet}.
	We denote by $\Sigma^*$ the set of finite words over $\Sigma$ and by $\epsilon$ the empty word.
	For $u \in \Sigma^*$, we respectively write $\pfx(u)$ and $\sfx(u)$ for the sets of prefixes and suffixes of $u$.
	We also let $\infx(u) \defeq \{v \in \Sigma^* \st \exists x,y \in \Sigma^* : u = xvy\}$.
	For every language $L \subseteq \Sigma^*$, we additionally define $\pfx(L) \defeq \bigcup_{u \in L} \pfx(u)$, $\sfx(L) \defeq \bigcup_{u \in L} \sfx(u)$, and $\infx(L) \defeq \bigcup_{u \in L} \infx(u)$.

	For a nonempty word $u \in \Sigma^*$ and $1 \leq i \leq |u|$, we denote by $u[i]$ the $i$th letter of $u$.
	Given $u \in \Sigma^*$ and $\ell \geq 1$, we denote by $u^\ell$ the word obtained by concatenating $\ell$ copies of $u$.
	Moreover, given $L \subseteq \Sigma^*$, we define $\first(L) \defeq \{ u[1] \st u \in L \setminus \{\epsilon\}\}$ and $\last(L) \defeq \{ u[|u|] \st u \in L \setminus \{\epsilon\}\}$.
	For sets $L_1, L_2 \subseteq \Sigma^*$ of words, we let $L_1 \cdot L_2 \defeq \{u \cdot v \st u \in L_1, v \in L_2\}$.
	For tuples $(u_1, \ldots, u_m)$ and $(v_1, \ldots, v_m)$ of words, we let $(u_1, \ldots, u_m) \cdot (v_1, \ldots, v_m) \defeq (u_1 v_1, \ldots, u_m v_m)$.
	
	We define the function $\destutter : \Sigma^* \to \Sigma^*$ inductively.
	For all $\sigma \in \Sigma \cup \{\epsilon\}$, let $\destutter(\sigma) \defeq \sigma$.
	For all $u \in \Sigma^*$ such that $u = \sigma_1 \sigma_2 v$ for some $\sigma_1,\sigma_2 \in \Sigma$ and $v \in \Sigma^*$, we define it as follows:
	\begin{equation*}
		\destutter(u) \defeq
		\begin{cases}
			\destutter(\sigma_2 v) & \text{if } \sigma_1 = \sigma_2 \\
			\sigma_1 \cdot \destutter(\sigma_2 v) & \text{otherwise}
		\end{cases}
	\end{equation*}

	For a set $L \subseteq \Sigma^*$ of finite words, we define $\destutter(L) \defeq \{\destutter(u) \st u \in L\}$.

	We extend $\destutter$ to tuples of words in a synchronized manner.
	For every tuple $(u_1, \ldots, u_m)$ of words of the same length at most $1$, let $\destutter(u_1, \ldots, u_m) \defeq (u_1, \ldots, u_m)$.
	Given $(u_1, \ldots, u_m) = (\sigma_{1,1}\sigma_{1,2}v_1, \ldots, \sigma_{m,1}\sigma_{m,2}v_m)$ with the words of the same length at least $2$, we define $\destutter(u_1, \ldots, u_m)$ as follows:
	\begin{align*}
		&\destutter(u_1, \ldots, u_m) \defeq \\
		&\begin{cases}
			\destutter(\sigma_{1,2} v_1, \ldots, \sigma_{m,2} v_m) \text{ if $\sigma_{i,1} = \sigma_{i,2}$ for}\\ 
			\hspace{14em}\text{all $1 \leq i \leq m$} \\
			(\sigma_{1,1}, \ldots, \sigma_{m,1}) \cdot \destutter(\sigma_{1,2} v_1, \ldots, \sigma_{m,2} v_m) \\
			\hspace{15.5em}\text{otherwise}
		\end{cases}
	\end{align*}
	
	Moreover, given an integer $k \geq 0$, we define $\stutter_k : \Sigma^* \to 2^{\Sigma^*}$ as $\stutter_k(u) \defeq \{v \in \Sigma^* \st |v| = k \land \destutter(v) = \destutter(u)\}$ if $k \geq |\destutter(u)|$, and $\stutter_k(u) \defeq \emptyset$ otherwise.
	
	\paragraph*{Signal Temporal Logic (STL) \cite{MalerN13}.}
	A \emph{temporal domain} is an interval $\mathbb{T} \subseteq \R_{\geq 0}$.
	A function $x : \mathbb{T} \to \R$ is \emph{piecewise constant} if $\mathbb{T}$ can be partitioned into countably many maximal intervals on each of which $x$ is constant.
	A piecewise-constant function $x$ has \emph{finite variability} if $\mathbb{T}$ can be partitioned into finitely many maximal intervals on each of which $x$ is constant; we call these intervals the \emph{constant pieces} of $x$.
	A \emph{signal} is a piecewise-constant function with finite variability.
	We write $\pi_\mathbb{T}(x)$ for the finite word that lists the values of the constant pieces from left to right, and define $\operatorname{var}_\mathbb{T}(x) \defeq |\pi_\mathbb{T}(x)| - 1$.
	If $x$ has a larger temporal domain, $\pi_\mathbb{T}(x)$ denotes the behavior of its restriction to $\mathbb{T}$.
	For every signal $x$ and every $t \in \mathbb{T}$ with $t > 0$, we respectively write $x(t^-)$ and $x(t^+)$ for its value immediately before and after $t$, i.e., $x(t^-) \defeq \lim_{s \to t^-} x(s)$ and $x(t^+) \defeq \lim_{s \to t^+} x(s)$.
	In particular, if $\mathbb{T} = [0,d)$, we write $x(d^-)$ for the value of its final constant piece.
	Note that the word $\pi_\mathbb{T}(x)$ is nonempty and destuttered.

	A signal $x : \mathbb{T} \to \R$ is \emph{right-continuous} if $\lim_{s \to t^+} x(s) = x(t)$ whenever the right limit is defined.
	An \emph{input signal} of duration $d \in \R_{>0}$ is a right-continuous signal with temporal domain $\mathbb{T} = [0,d)$.
	Let $x : [0,d) \to \R$ be an input signal.
	An \emph{event} of the input signal $x$ is a pair $(t, x(t))$ where $t \in [0,d)$.
	An \emph{edge} of the input signal $x$ is an event $(t, x(t))$ with $t \in (0,d)$ such that $x(t^-) \neq x(t^+)$.
	In particular, an edge is \emph{rising} if $x(t^-) < x(t^+)$ and it is \emph{falling} otherwise.
	An input signal $x : [0,d) \to \R$ can be represented finitely by its initial value and edges: if $x$ has $m$ edges, then $x = (t_0, v_0) (t_1, v_1) \ldots (t_m, v_m)$ such that $t_0 = 0$, $v_0 = x(0)$, $t_{i-1} < t_i$, and $(t_i, v_i)$ is an edge of $x$ for all $1 \leq i \leq m$.
	
	Let $\AP$ be a set of \emph{atomic propositions}.
	The syntax of \emph{future STL} is given by the grammar $\varphi \defeq \true ~|~ p ~|~ \lnot \varphi ~|~ \varphi \land \varphi ~|~ \varphi \until_I \varphi$ and that of \emph{past STL} is given by $\varphi \defeq \true ~|~ p ~|~ \lnot \varphi ~|~ \varphi \land \varphi ~|~ \varphi \since_I \varphi$ where $p \in \AP$ and $I \subseteq \R_{\geq 0}$ is an interval.
	Below, when we speak of an STL formula, we mean either a future STL or a past STL formula.

	A \emph{trace} of duration $d \in \R_{>0}$ is a finite vector $w = (x_1, \ldots, x_n)$ of signals with the common temporal domain $[0,d)$.
	Each atomic proposition $p \in \AP$ is interpreted by a function $f_p : \R^n \to \R$.
	For a trace $w = (x_1, \ldots, x_n)$ and a time point $t$, we evaluate $p$ on $w$ at time $t$ by the predicate $f_p(x_1(t), \ldots, x_n(t)) > 0$.
	Given intervals $I,J \subseteq \R_{\geq 0}$, we define $I \oplus J \defeq \{i + j \st i \in I , j \in J\}$ and $I \ominus J \defeq \{i - j \st i \in I , j \in J\}$, where we simply write $t$ for the singleton set $\{t\}$.
	
	We recall the finite-trace qualitative semantics of STL defined over $\B$.
	Let $w = (x_1, \ldots, x_n)$ be a trace of duration $d$.
	Let $\varphi_1, \varphi_2$ be STL formulas and let $t \in [0,d)$.
	In general, $(w,t) \models \true$.

	\begin{align*}
		(w,t) \models p \iff & f_p(x_1(t), \ldots, x_n(t)) > 0 \\
		(w,t) \models \lnot \varphi_1 \iff & (w,t) \not\models \varphi_1 \\
		(w,t) \models \varphi_1 \land \varphi_2 \iff & (w,t) \models \varphi_1 \land (w,t) \models \varphi_2 \\
		(w,t) \models \varphi_1 \until_I \varphi_2 \iff & \exists t' \in (t \oplus I) \cap [0,d) :  \\
		& (w,t') \models \varphi_2 \land \forall t'' \in [t, t') : \\
		& (w,t'') \models \varphi_1\\
		(w,t) \models \varphi_1 \since_I \varphi_2 \iff & \exists t' \in (t \ominus I) \cap [0,d) :  \\
		& (w,t') \models \varphi_2 \land \forall t'' \in (t', t] : \\
		& (w,t'') \models \varphi_1
	\end{align*}
	
	For a trace $w$ and an STL formula $\varphi$, the \emph{satisfaction signal} of $\varphi$ on $w$ is the Boolean signal $y_{\varphi,w} : [0,d) \to \B$ defined by
	\begin{equation*}
		y_{\varphi,w}(t) = 1 \quad \Longleftrightarrow \quad (w,t) \models \varphi.
	\end{equation*}
	Finite variability of $w$ implies finite variability of $y_{\varphi,w}$ by structural induction on $\varphi$.
	Note that a satisfaction signal need not be right-continuous.
	For a future STL formula $\varphi$, we write $w \models \varphi$ if $y_{\varphi,w}(0) = 1$.
	For a past STL formula $\varphi$, we write $w \models \varphi$ if $y_{\varphi,w}(d^-) = 1$.
	We additionally use the following abbreviations:
	$\false \defeq \lnot \true$,
	$\varphi_1 \lor \varphi_2 \defeq \lnot (\lnot \varphi_1 \land \lnot \varphi_2)$,
	$\LTLf_I \varphi \defeq \true \until_I \varphi$,
	$\LTLg_I \varphi \defeq \lnot \LTLf_I \lnot \varphi$,
	$\LTLdiamondminus_I \varphi \defeq \true \since_I \varphi$, and
	$\LTLsquareminus_I \varphi \defeq \lnot \LTLdiamondminus_I \lnot \varphi$.
	Moreover, the untimed temporal operators are defined through their timed counterparts on the interval $[0,\infty)$.
	An STL formula is \emph{untimed} if every temporal operator in its syntax uses the interval $[0,\infty)$, and it is \emph{timed} otherwise.
	
	Note that future STL formulas whose timed operators are bounded (i.e., use of $\until_I$ is restricted to intervals $I$ with a real-valued right endpoint) can be translated to past STL formulas, which can be efficiently monitored online~\cite{MalerNP07,JaksicBGKNN15,Gol18}.

	\paragraph*{Distributed Semantics of STL \cite{MomtazAB23}.}
	We consider an asynchronous and loosely-coupled message-passing system of $n \geq 2$ reliable agents producing a set of input signals $x_1, \ldots, x_n$, where for some $d \in \R_{> 0}$ we have $x_i : [0,d) \to \R$ for all $1 \leq i \leq n$.
	The agents do not share memory or a global clock.
	We assume that the system is \emph{partially synchronous}: the agents use a clock synchronization algorithm that guarantees a bound $\varepsilon \in \R_{> 0}$ on pairwise clock skew, so the readings of any two clocks at the same physical instant differ by less than $\varepsilon$.
	To formalize statements, we introduce an unobservable semantic reference clock $c_0 : [0,d) \to [0,d)$ defined by $c_0(T) \defeq T$; it is not associated with or observed by any agent.
	For local time values, we use the lowercase letters $t$ and $s$.
	For an input signal $x_i$, we denote by $V_i$ the set of its events.
	Although we write events of $x_i$ as pairs $(t, x_i(t))$, we regard $V_1, \ldots, V_n$ as pairwise disjoint copies.
	Thus, the signal identity is implicit in the choice of $V_i$.
	We represent a possible correspondence between reference time and local clock readings by a clock family $c = (c_0, \ldots, c_n)$, where $c_i(T)$ is the reading of the $i$th agent's local clock at reference time $T$ for every $1 \leq i \leq n$.
	Such a clock family is \emph{admissible} if every $c_i : [0,d) \to [0,d)$ is a strictly increasing bijection and $|c_i(T) - c_j(T)| < \varepsilon$ for all $0 \leq i,j \leq n$ and $T \in [0,d)$.
	
	\begin{definition} \label{defn:hb}
		A \emph{distributed signal} is a pair $(S, {\hb})$, where $S = (x_1, \ldots, x_n)$ is a vector of input signals and the happened-before relation ${\hb}$ is the least strict partial order on $V = \biguplus_{i=1}^n V_i$ satisfying the following conditions:
		(1) For every agent, the events of its input signal are totally ordered, i.e., for all $1 \leq i \leq n$ and all $(t, x_i(t)), (t', x_i(t')) \in V_i$, if $t < t'$ then $(t, x_i(t)) \hb (t', x_i(t'))$.
		(2) Every pair of events whose time\-stamps are at least $\varepsilon$ apart is totally ordered, i.e., for all $1 \leq i,j \leq n$ and all $(t, x_i(t)) \in V_i$ and $(t', x_j(t')) \in V_j$, if $t + \varepsilon \leq t'$ then $(t, x_i(t)) \hb (t', x_j(t'))$.
	\end{definition}
	
	The notion of \emph{consistent cut} captures possible global states.
	
	\begin{definition}
		Let $(S, {\hb})$ be a distributed signal of $n$ input signals, and $V = \biguplus_{i = 1}^{n} V_i$ be the set of its events.
		A set $C \subseteq V$ is a \emph{consistent cut} iff for every event in $C$, all events that happened before it also belong to $C$, i.e., for all $e, e' \in V$, if $e \in C$ and $e' \hb e$, then $e' \in C$.
	\end{definition}

	Given a set $C \subseteq V$ that has a maximal event on each input signal (i.e., for each input signal $x_i$, an event in $C \cap V_i$ whose timestamp is greatest among the events in $C \cap V_i$), its \emph{frontier} $\fr(C) \subseteq C$ is the set consisting of these maximal events, i.e., $\fr(C) \defeq \bigcup_{i = 1}^{n} \{ (t, x_i(t)) \in V_i \cap C \st \forall t' > t : (t', x_i(t')) \notin V_i \cap C \}$.
	
	\begin{definition}
		Let $c = (c_0, \ldots, c_n)$ be an admissible clock family.
		For every $T \in [0,d)$, let $C_c(T) \defeq \bigcup_{i=1}^n \{(t,x_i(t)) \in V_i \mid t \leq c_i(T)\}$.
		We call $c$ \emph{consistent} iff $C_c(T)$ is a consistent cut for every $T \in [0,d)$.
		For a consistent $c$, the \emph{consistent cut flow} induced by $c$ is the function $\ccf_c : [0,d) \to 2^V$ defined by $\ccf_c(T) \defeq \fr(C_c(T)) = \{(c_i(T),x_i(c_i(T))) \mid 1 \leq i \leq n\}$.
	\end{definition}

	By mapping the parameter $T$ to the frontier of a consistent cut, a consistent cut flow captures a possible evolution of the system's global state over time.
	Each such global state is represented by a consistent cut.
	\Cref{fig:distsig} shows three consistent cuts that can occur along a consistent cut flow and a cut that violates the happened-before relation.
	
	\begin{figure}
		\centering
		\begin{tikzpicture}[scale=0.75]
			\draw[very thick, ->] (0,0) node[below] {$x_1$} -- (0,3.5) ;
			\draw[very thick, ->] (2,0) node[below] {$x_2$} -- (2,3.5) ;
			
			\draw[admNeutralLine] (-1,0.45) -- (3,0.65) node[right, below right, black] {C$_1$};
			\draw[admNeutralLine] (-1,1.8) -- (3,1.0) node[right, black] {C$_2$};
			\draw[admNeutralLine] (-1,2.95) -- (3,2.75) node[right, black] {C$_3$};
			
			\draw[admAccentRed, dashed] (-1,0.5) -- (3,3.25) node[midway, above, black] {C$'$};
			
			\node[left] at (0,0.3) {{\scriptsize 0.5}};
			\node[left] at (0,1.5) {{\scriptsize 1.6}};
			\node[left] at (0,3.1) {{\scriptsize 2.9}};
			\draw[fill=black] (0,0.5) circle (1pt);
			\draw[fill=black] (0,1.6) circle (1pt);
			\draw[fill=black] (0,2.9) circle (1pt);
			
			\node[right] at (2,0.4) {{\scriptsize 0.6}};
			\node[right] at (2,1.3) {{\scriptsize 1.2}};
			\node[right] at (2,2.4) {{\scriptsize 2.5}};
			\node[right] at (2,3.0) {{\scriptsize 2.8}};	
			\draw[fill=black] (2,0.6) circle (1pt);
			\draw[fill=black] (2,1.2) circle (1pt);
			\draw[fill=black] (2,2.5) circle (1pt);
			\draw[fill=black] (2,2.8) circle (1pt);
		\end{tikzpicture}
		\caption{A distributed signal with clock-skew bound $\varepsilon = 0.5$.
		The solid lines show three consistent cuts $C_1$, $C_2$, and $C_3$ that can occur along a consistent cut flow.
		The dashed cut $C'$ is not consistent since it contains $(2.5,x_2(2.5))$ but not $(1.6,x_1(1.6))$, although $1.6 + \varepsilon = 2.1 \leq 2.5$ and hence $(1.6,x_1(1.6)) \hb (2.5,x_2(2.5))$.}
		\label{fig:distsig}
	\end{figure}

	Let $(S,{\hb})$ be a distributed signal of $n$ input signals $x_1, \ldots, x_n$.
	Every consistent admissible clock family $c$ induces the \emph{synchronous trace} (a.k.a. \emph{retiming}) $w_c \defeq (x_1 \circ c_1, \ldots, x_n \circ c_n)$.
	The set of traces of $(S,{\hb})$ is therefore
	\begin{equation*}
		\tr(S,{\hb}) \defeq
		\{w_c \mid \text{$c$ is consistent and admissible}\}.
	\end{equation*}
	The set $\tr(S,{\hb})$ is nonempty because $c_0 = \ldots = c_n$ is a consistent admissible clock family that induces $S$.
	Moreover, an edge of $x_i$ at local time $t$ occurs exactly once in $x_i \circ c_i$, at reference time $T = c_i^{-1}(t)$.
	Since $c_i^{-1}$ is strictly increasing, these occurrence times preserve the source order.
	Furthermore, $|T-t| = |c_0(T)-c_i(T)| < \varepsilon$.
	
	We define the satisfaction of an STL formula $\varphi$ by a distributed signal $(S,{\hb})$ over a three-valued domain $\{\top, \bot, {?}\}$.
	Notice that we quantify universally over traces for both satisfaction and violation.
	\begin{equation*}
		[(S,{\hb}) \models \varphi] \defeq
		\begin{cases}
			\top \text{ if } \forall w \in \tr(S,{\hb}) : w \models \varphi \\
			\bot \text{ if } \forall w \in \tr(S,{\hb}) : w \models \lnot\varphi \\
			\,?\, \text{ otherwise}
		\end{cases}
	\end{equation*}

	\section{Monitoring by Two-Step Overapproximation}
	\label{sec:semantics}

	Exact distributed monitoring works directly on the trace set $\tr(S,{\hb})$ (\cref{sec:preliminaries}).
	Our monitor, instead, overapproximates the set in two steps: first at the level of traces, and then at the level of satisfaction signals.

	In the first step (\cref{sec:approach}), we construct a larger trace set $\tr^+(S,{\hb})$.
	The key observation is that partial synchrony leaves only edge timing uncertain.
	Any consistent admissible clock family only shifts each edge by less than $\varepsilon$, effectively restricting it to an \emph{uncertainty region} around its recorded timestamp.
	Accordingly, the construction partitions the temporal domain into \emph{canonical segments} at the region endpoints and computes, for each component signal and each segment, a finite set of \emph{value expressions}, i.e., words representing its possible behaviors.
	The set $\tr^+(S,{\hb})$ then consists of all traces for which the word induced by each component on each segment belongs to the corresponding set.
	Every trace in $\tr(S,{\hb})$ satisfies these local conditions, since each is necessary for being induced by a consistent admissible clock family.
	They are not jointly sufficient, however, so $\tr^+(S,{\hb})$ may also contain spurious traces that no such clock family induces.

	In the second step (\cref{sec:algorithm}), we compute, for every subformula $\psi$ and every canonical segment $I$, the \emph{symbolic output} $\llbracket (S,{\hb}), I \models \psi \rrbracket$: a finite set of words describing the possible value expressions of the satisfaction signal of $\psi$ on $I$.
	The computation lifts the semantics of STL from signals to these word sets.
	It proceeds bottom-up over the formula, with one rule per operator, and never constructs $\tr^+(S,{\hb})$ itself.
	For an atomic proposition, the evaluator aligns component value expressions from the first step and applies the numerical predicate pointwise.
	Boolean and untimed temporal operators combine the value expressions of their operands within the segment, considering all the possible alignments, and exchange at most a single bit with the adjacent segment.
	A timed operator, by contrast, may need information from several segments because its timing interval can extend beyond $I$.
	Its rule combines the symbolic outputs of the subformulas on all segments that intersect the relevant time range, using only the parts that lie inside that range, while accounting for how this range moves as the evaluation time varies over $I$.

	Together, the two steps yield the following chain of inclusions:
	\[
		\begin{gathered}
			\{\pi_I(y_{\psi,w}) \st w\in\tr(S,{\hb})\}\\[-0.2ex]
			\subseteq\\[-0.2ex]
			\{\pi_I(y_{\psi,w}) \st w\in\tr^+(S,{\hb})\}\\[-0.2ex]
			\subseteq\\[-0.2ex]
			\llbracket (S,{\hb}), I \models \psi \rrbracket.
		\end{gathered}
	\]
	We will prove the first inclusion in \cref{cl:trsound} and the second in \cref{cl:evalsound}.

	The monitor obtains its verdict from the first letter of every word in the symbolic output on the first canonical segment for a future formula, and the last letter of every word on the last canonical segment for a past formula.
	The inclusions above ensure that the relevant symbolic output contains the value expression induced by every trace in $\tr(S,{\hb})$.
	If all the selected letters are $1$, every trace satisfies $\varphi$, and the monitor returns $\top$.
	If all of them are $0$, every trace violates $\varphi$, and the monitor returns $\bot$.
	Otherwise, it returns ${?}$.
	\Cref{cl:algo} proves that these verdicts are sound with respect to the standard verdict $[(S,{\hb}) \models \varphi]$.

	\section{Overapproximation of Synchronous Traces} 
	\label{sec:approach}
	
	Given a distributed signal $(S,{\hb})$, we construct an overapproximation $\tr^+(S,{\hb})$ of $\tr(S,{\hb})$.
	We first \emph{partition} $[0,d)$ at the endpoints of edge uncertainty regions.
	We then use finite \emph{value expressions} to describe the possible behavior of each component on each canonical segment.
	Finally, we define $\tr^+(S,{\hb})$ from these local segment conditions and prove that they admit every trace in $\tr(S,{\hb})$.

	\paragraph*{Canonical Segmentation.}
	Let $(t,x_i(t))$ be an edge of a component input signal $x_i$.
	For each consistent admissible clock family $c$, this edge occurs in $w_c$ at reference time $T = c_i^{-1}(t)$.
	Since $|T-t| < \varepsilon$, the occurrence lies in the interval $(\theta_{\mathrm{lo}}(x_i,t),\theta_{\mathrm{hi}}(x_i,t))$, where $\theta_{\mathrm{lo}}(x_i,t) \defeq \max(0,t-\varepsilon)$ and $\theta_{\mathrm{hi}}(x_i,t) \defeq \min(d,t+\varepsilon)$.
	We call this interval the \emph{uncertainty region} of the edge.

	A \emph{segmentation} of $[0,d)$ is the partition induced by finitely many points $0 = t_1 < \ldots < t_{k+1} = d$.
	Its segments are $I_j \defeq [t_j,t_{j+1})$ for $1 \leq j\leq k$.
	By extension, a segmentation of signals with the common temporal domain $[0,d)$ is a segmentation of $[0,d)$.

	Let $(S,{\hb})$ be a distributed signal of $n$ signals.
	The \emph{canonical segmentation} $G_S$ of $(S,{\hb})$ is the segmentation of the common temporal domain whose endpoints are exactly $0$, $d$, and the uncertainty-region endpoints.
	Formally, for each signal $x_i$, where $1 \leq i \leq n$, let $F_i \defeq \{ \theta_{\text{lo}}(x_i,t), \theta_{\text{hi}}(x_i,t) \mid (t,x_i(t)) \text{ is an edge of } x_i \}$.
	Let $F \defeq \{0, d\} \cup \bigcup_{i = 1}^{n} F_i$, and let $s_1 < \ldots < s_{|F|}$ be the increasing enumeration of $F$.
	Then, the canonical segmentation is $G_S \defeq \{I_1, \ldots, I_{|F| - 1}\}$ where $I_j \defeq [s_j, s_{j+1})$ for all $1 \leq j < |F|$.
	We show an example in \cref{fig:csve}a.

	\begin{figure*}
		\centering
		\begin{subfigure}[c]{.45\textwidth}
			\centering
			\begin{tikzpicture}[scale=0.75]
				\scriptsize
				\node at (-1.5,0) {\textbf{(a)}};
				\footnotesize 
				% Plot for x1(t)
				\begin{scope}
					% Axes
					\draw[->] (0,0) -- (8.5,0) node[right] {$t$};
					\draw[->] (0,0) -- (0,1.5) node[above] {$x_1(t)$};
					
					% Red shaded areas
					\fill[black!15, fill opacity=.45] (0,1.1) rectangle (4,0.9);
					\fill[black!15, fill opacity=.45] (3,0.1) rectangle (7,-0.1);
					
					% Red dashed lines
					\draw[admAccentBlue, thick, dashed] (0,0) -- (2,0) -- (2,1) -- (5,1) -- (5,0) -- (8,0);
					
					% Red dots
					\fill[admAccentBlue] (0,0) circle (2pt);
					\fill[admAccentBlue] (2,1) circle (2pt);
					\fill[admAccentBlue] (5,0) circle (2pt);
				\end{scope}
				
				% Plot for x2(t)
				\begin{scope}[shift={(0,-2.5)}]
					% Axes
					\draw[->] (0,0) -- (8.5,0) node[right] {$t$};
					\draw[->] (0,0) -- (0,1.5) node[above] {$x_2(t)$};
					
					% Red shaded areas
					\fill[black!15, fill opacity=.45] (1,1.1) rectangle (5,0.9);
					\fill[black!15, fill opacity=.45] (4,0.1) rectangle (8,-0.1);
					
					% Red dashed lines
					\draw[admAccentOchre, thick, dashed] (0,0) -- (3,0) -- (3,1) -- (6,1) -- (6,0) -- (8,0);
					
					% Red dots
					\fill[admAccentOchre] (0,0) circle (2pt);
					\fill[admAccentOchre] (3,1) circle (2pt);
					\fill[admAccentOchre] (6,0) circle (2pt);
				\end{scope}
				
				% GS boxes
				\begin{scope}[shift={(0,-4)}]
					\draw (0,0) rectangle (1,1);
					\node at (0.5,0.5) {{\scriptsize $[0,1)$}};
					\draw (1,0) rectangle (3,1);
					\node at (2,0.5) {{\scriptsize $[1,3)$}};
					\draw (3,0) rectangle (4,1);
					\node at (3.5,0.5) {{\scriptsize $[3,4)$}};
					\draw (4,0) rectangle (5,1);
					\node at (4.5,0.5) {{\scriptsize $[4,5)$}};
					\draw (5,0) rectangle (7,1);
					\node at (6,0.5) {{\scriptsize $[5,7)$}};
					\draw (7,0) rectangle (8,1);
					\node at (7.5,0.5) {{\scriptsize $[7,8)$}};
				\end{scope}
			\end{tikzpicture}
		\end{subfigure}
		\hspace{2em}
		\begin{subfigure}[c]{.45\textwidth}
			\centering
			\begin{tikzpicture}[scale=0.65]
				\scriptsize
				% Part (a)
				\node at (-1.5,0) {\textbf{(b)}};
				\draw[thick] (0,0) -- (8,0);
				\foreach \x in {0,1,3,4,5,7,8}
				\draw (\x,0) node[above] {\x};
				\draw (-0.5,0) node[above] {$t$:};
				
				\draw[fill=admAccentBlue!7] (-1,-2) rectangle (0,0) node[midway] {$x_1$};
				\draw[fill=white] (0,-1) rectangle (4,0) node[midway] {$01$};
				\draw[fill=white] (3,-2) rectangle (7,-1) node[midway] {$10$};
				\draw[fill=white] (7,-2) rectangle (8,0) node[midway] {$0$};
				
				\draw[fill=admAccentOchre!7] (-1,-4) rectangle (0,-2) node[midway] {$x_2$};
				\draw[fill=white] (0,-4) rectangle (1,-2) node[midway] {$0$};
				\draw[fill=white] (1,-3) rectangle (5,-2) node[midway] {$01$};
				\draw[fill=white] (4,-4) rectangle (8,-3) node[midway] {$10$};
				
				\draw (0,-4) -- (8,-4);
				\draw (8,-4) -- (8,0);
				
				% Part (b)
				\node at (-1.5,-5) {\textbf{(c)}};
				\draw[thick] (0,-5) -- (8,-5);
				\foreach \x in {0,1,3,4,5,7,8}
				\draw (\x,-5) node[above] {\x};
				\draw (-0.5,-5) node[above] {$t$:};
				
				\draw[fill=admAccentBlue!7] (-1,-7) rectangle (0,-5) node[midway] {$x_1$};
				\draw[fill=white] (0,-7) rectangle (1,-5) node[midway] { \fontsize{8}{8}\selectfont \begin{tabular}{c}\{0,\\ 01\}\\\end{tabular}};
				\draw[fill=white] (1,-7) rectangle (3,-5) node[midway] {\fontsize{8}{8}\selectfont\begin{tabular}{c}\{0,\\ 01,\\ 1\}\\\end{tabular}};
				\draw[fill=white] (3,-7) rectangle (4,-5) node[midway] {\fontsize{8}{8}\selectfont{\begin{tabular}{c}\{01,\\ 010,\\ 1,\\ 10\}\\\end{tabular}}};
				\draw[fill=white] (4,-7) rectangle (5,-5) node[midway] {\fontsize{8}{8}\selectfont\begin{tabular}{c}\{1,\\ 10,\\ 0\}\\\end{tabular}};
				\draw[fill=white] (5,-7) rectangle (7,-5) node[midway] {{\fontsize{8}{8}\selectfont\begin{tabular}{c}\{10,\\ 0\}\\\end{tabular}}};
				\draw[fill=white] (7,-7) rectangle (8,-5) node[midway] {{\fontsize{8}{8}\selectfont\begin{tabular}{c}\{0\}\\\end{tabular}}};
				
				\draw[fill=admAccentOchre!7] (-1,-9) rectangle (0,-7) node[midway] {$x_2$};
				\draw[fill=white] (0,-9) rectangle (1,-7) node[midway] {{\fontsize{8}{8}\selectfont\begin{tabular}{c}\{0\}\\\end{tabular}}};
				\draw[fill=white] (1,-9) rectangle (3,-7) node[midway] {\fontsize{8}{8}\selectfont\begin{tabular}{c}\{0,\\ 01\}\\\end{tabular}};
				\draw[fill=white] (3,-9) rectangle (4,-7) node[midway] {\fontsize{8}{8}\selectfont\begin{tabular}{c}\{0,\\ 01,\\ 1\}\\\end{tabular}};
				\draw[fill=white] (4,-9) rectangle (5,-7) node[midway] {\fontsize{8}{8}\selectfont{\begin{tabular}{c}\{01,\\ 010,\\ 1,\\ 10\}\\\end{tabular}}};
				\draw[fill=white] (5,-9) rectangle (7,-7) node[midway] {\fontsize{8}{8}\selectfont\begin{tabular}{c}\{1,\\ 10,\\ 0\}\\\end{tabular}};
				\draw[fill=white] (7,-9) rectangle (8,-7) node[midway] {{\fontsize{8}{8}\selectfont\begin{tabular}{c}\{10,\\ 0\}\\\end{tabular}}};
			\end{tikzpicture}
		\end{subfigure}
		\caption{\textbf{(a)} A distributed signal $(S,{\hb})$ of duration $d = 8$ and clock-skew bound $\varepsilon = 2$ with the components $x_1$ (top) and $x_2$ (bottom).
		The edges are marked with solid balls and their uncertainty regions are given as shaded boxes around the edges.
		The resulting canonical segmentation $G_S$ is shown below the graphical representation of the signals.
		\textbf{(b)} The uncertainty regions of $(S,{\hb})$ and the corresponding value expressions.
		\textbf{(c)} The tabular representation of the function $\gamma$ for $(S,{\hb})$, e.g., $\gamma(x_1, [3,4)) = \destutter(\sfx(01) \cdot \pfx(10)) \setminus \{\epsilon\} = \{ 01, 010, 1, 10\}$.\label{fig:csve}}
	\end{figure*}
	
	\paragraph*{Value Expressions.}
	Consider a distributed signal $S$ and let $\Sigma$ be the finite set of values that occur in its component signals.
	A \emph{value expression} is a word in $\Sigma^*$.
	For example, a Boolean rising edge has the value expression $01$.
	Given a component input signal $x$ and an edge $(t,x(t))$, the value expression corresponding to its uncertainty region is $v_{x,t} \defeq x(t^-) \cdot x(t^+)$.
	We omit the concatenation symbol $\cdot$ when the letters are clear from context.

	Notice that uncertainty regions may overlap and the canonical segmentation may split an uncertainty region into multiple segments.
	Consider a signal $x$ that has a rising edge with the uncertainty region $(1,5)$ and a falling edge with $(4,8)$.
	The corresponding value expressions are respectively $v_1 = 01$ and $v_2 = 10$.
	The behavior of $x$ in the interval $[1,4)$ can be expressed as $\pfx(v_1) \setminus \{\varepsilon\}$, encoding whether the rising edge has happened yet.
	Similarly, the behavior in $[4,5)$ is given by $\sfx(v_1) \cdot \pfx(v_2) \setminus \{\varepsilon\}$, which captures whether the edges occur in this interval (by prefixing and suffixing) and the fact that the rising edge happens before the falling edge (by concatenation).

	Formally, given a distributed signal $(S,{\hb})$, we define a function $\gamma$ that maps each component signal $x$ of $S$ and each segment	$I \in G_S$ to a set of potential behaviors of $x$ on $I$.
	Consider a component signal $x$ and let $R_1, \ldots, R_m$ be its uncertainty regions, listed from earliest to latest according to the timestamps of the corresponding edges.
	For each $1 \leq j \leq m$, write $R_j = (a_j,b_j)$ and let $v_j$ be the value expression of the same source edge.
	For $I = [s,s') \in G_S$, we define
	\[
		\mathcal L_j(x,I)\defeq
		\begin{cases}
			\{v_j\} & \text{if }s\leq a_j<b_j\leq s',\\
			\pfx(v_j) & \text{if }s\leq a_j<s'<b_j,\\
			\sfx(v_j) & \text{if }a_j<s<b_j\leq s',\\
			\infx(v_j) & \text{if }a_j<s<s'<b_j,\\
			\{\epsilon\} & \text{if }s'\leq a_j\text{ or }b_j\leq s.
		\end{cases}
	\]
	Note that the last case holds when $I \cap R_j = \emptyset$.

	If no uncertainty region of $x$ has a nonempty intersection with $I$, we define
	\[
		\gamma(x,I) \defeq \{x(s)\}.
	\]
	Otherwise, define
	\[
		\gamma(x,I) \defeq \destutter(\mathcal L_1(x,I) \cdot \ldots \cdot \mathcal L_m(x,I)) \setminus \{\epsilon\}.
	\]

	If $I$ does not intersect any uncertainty region of $x$, no source edge can be retimed into $I$; so, the only possible value expression is the one-letter word $x(s)$.
	Otherwise, $\gamma(x,I)$ accounts for all possible sequences of relevant source edges that may occur in $I$, while preserving their original order.
	As we will show in \cref{cl:trsound}, every value expression produced by a retiming belongs to $\gamma(x,I)$.
	However, $\gamma(x,I)$ may also contain additional value expressions because it chooses one behavior from each $\mathcal L_j(x,I)$ without checking whether all these behaviors can arise from the same retiming.

	\paragraph*{Segmentwise Trace Overapproximation.}
	Consider a distributed signal $(S,{\hb})$ of $n$ component signals $S = (x_1, \ldots, x_n)$, and let $G_S$ be its canonical segmentation.
	Let $w' = (x'_1,\ldots,x'_n)$ range over traces of duration $d$.
	We define
	\[
		\tr^+(S,{\hb}) \defeq
		\left\{
			w' \st
			\begin{array}{l}
				\pi_I(x'_i) \in \gamma(x_i,I) \text{ for all}\\
				1 \leq i \leq n \text{ and } I \in G_S
			\end{array}
		\right\}.
	\]

	Notice that the definition imposes the condition $\pi_I(x'_i) \in \gamma(x_i,I)$ separately for each component $i$ and each canonical segment $I$.
	As \Cref{ex:overapx} shows, this locality may lead to the same source edge being represented in several segments.

	\begin{example} \label{ex:overapx}
		Recall $(S,{\hb})$ and its $\gamma$ function from \cref{fig:csve}.
		Consider the synchronous trace $w \in \tr(S,{\hb})$ where the rising edges of both component signals occur at time 3 and the falling edges at time 5.
		This trace $w$ belongs to $\tr^+(S,{\hb})$ since for each $i \in \{1,2\}$ the value expression $1$ belongs to $\gamma(x_i,[3,4))$ and $\gamma(x_i,[4,5))$, while $0$ belongs to the remaining sets for $x_i$.

		Now, consider the trace $w' = (x'_1,x'_2)$ where each component signal is initially 0, rises at time 2, falls at time 3, rises again at time 3.5, and falls again at time 5.
		For each $i \in \{1,2\}$, its value expression is $01$ on $[1,3)$ and $[3,4)$, $1$ on $[4,5)$, and $0$ on the remaining segments.
		So, for every component $i$ and every canonical segment $I$, the value expression $\pi_I(x'_i)$ belongs to $\gamma(x_i,I)$, and thus $w' \in \tr^+(S,{\hb})$.
		However, $w' \notin \tr(S,{\hb})$ since every retiming preserves the number and order of the source edges on each component, while $w'$ represents the same edges in multiple segments.
	\end{example}

	Although these local conditions admit traces that do not arise from any single retiming, they are satisfied by every retiming: each retimed source edge lies within its uncertainty region, and the retiming preserves the order of the source edges.
	Therefore, for every component $x_i$ and every canonical segment $I$, the value expression of the retimed component on $I$ belongs to $\gamma(x_i,I)$.
	We formalize this argument and prove the first inclusion in \cref{sec:semantics}.

	\begin{lemma} \label{cl:trsound}
		For every distributed signal $(S,{\hb})$, we have $\tr(S,{\hb})\subseteq\tr^+(S,{\hb})$.
	\end{lemma}
	\begin{proof}[\normalsize Proof.]
		\normalsize
		Let $w_c = (x_1^c, \ldots, x_n^c) \in \tr(S,{\hb})$ be induced by a consistent admissible clock family $c = (c_0, \ldots, c_n)$, where $x_i^c \defeq x_i \circ c_i$.
		Consider a component $i$ and a canonical segment $I = [s,s')$.
		Let $e_1, \ldots, e_m$ be the source edges of $x_i$ in the order they appear in $x_i$.
		For each $j$, let $t_j$ be the timestamp of $e_j$ and let $R_j = (a_j,b_j)$ be its uncertainty region.
		Let us write $v_j = \sigma_j^- \sigma_j^+$ for the value expression of $e_j$.
		Let $T_j \defeq c_i^{-1}(t_j)$, i.e., the edge $e_j$ occurs at time $T_j$ in the retimed signal $x_i^c$.
		Let $L_j \defeq \mathcal L_j(x_i,I)$.

		By the edge-retiming property, we have $T_1 < \ldots < T_m$ and $T_j \in R_j$ for every $j$.
		A region $R_j = (a_j,b_j)$ has a nonempty intersection with $I = [s,s')$ exactly when $a_j < s'$ and $b_j > s$.
		Since $t_1 < \ldots < t_m$, both $(a_j)_{1 \leq j \leq m}$ and $(b_j)_{1 \leq j \leq m}$ are nondecreasing.
		So, if $R_j$ and $R_\ell$ have a nonempty intersection with $I$ satisfying $j < k < \ell$, then $a_k \leq a_\ell < s'$ and $b_k \geq b_j > s$.
		Therefore, $R_k$ also has a nonempty intersection with $I$, so the relevant indices form a consecutive block.
		We prove that $\pi_I(x_i^c) \in \gamma(x_i,I)$.

		Suppose first that no uncertainty region has a nonempty intersection with $I$.
		Each $R_j$ then lies completely before or completely after $I$, and both $t_j$ and $T_j$ lie in $R_j$.
		So, $x_i$ and $x_i^c$ have crossed the same source edges by time $s$, and no occurrence lies in $I$.
		Therefore, $\pi_I(x_i^c) = x_i(s) \in \gamma(x_i,I)$.

		Now, suppose that at least one uncertainty region has a nonempty intersection with $I$.
		For every $j$, if $T_j \in (s,s')$, then $T_j \in R_j \cap I$, so $L_j$ contains $v_j$.
		If $T_j \notin (s,s')$, then $L_j$ contains $\epsilon$, because the only case without $\epsilon$ is $L_j = \{v_j\}$, which would imply $T_j \in (s,s')$.

		Suppose that at least one occurrence lies in $(s,s')$.
		Since the times $T_j$ are strictly increasing, the corresponding indices form a consecutive block $p, \ldots, q$.
		Choose $u_j = v_j$ for $p \leq j\leq q$ and choose $u_j = \epsilon$ otherwise.
		The preceding observations give $u_j \in L_j$ for every $j$.
		Consecutive source edges satisfy $\sigma_j^+ = \sigma_{j+1}^-$, so $\destutter(u_1 \ldots u_m) = \sigma_p^- \sigma_p^+ \ldots \sigma_q^+$.
		Since $T_p$ is the first occurrence strictly after $s$, the first letter is $x_i^c(s)$ and the remaining letters are the values after the occurrences in $(s,s')$.
		Hence, $\destutter(u_1 \ldots u_m) = \pi_I(x_i^c)$.

		Otherwise, no occurrence lies in $(s,s')$, so $x_i^c$ is constant on $I$.
		Let $z \defeq x_i^c(s)$, and let $e_\ell, \ldots, e_r$ be the consecutive block of source edges whose regions have a nonempty intersection with $I$.
		Every source edge before $e_\ell$ has its occurrence before $s$, and every source edge after $e_r$ has its occurrence after $s'$.
		If some relevant occurrence is at or before $s$, let $k$ be the largest corresponding index.
		Every later relevant occurrence is at or after $s'$, so $e_k$ is the last source edge that has occurred by $s$, and $z = \sigma_k^+$.
		Since $R_k$ has a nonempty intersection with $I$ and contains $T_k \leq s$, we have $a_k < s < b_k$.
		Thus, $L_k$ is a suffix or infix language and contains $z$.
		In this case, choose $u_k = z$.
		If no relevant occurrence is at or before $s$, then every relevant occurrence is at or after $s'$.
		All source edges before $e_\ell$ have occurred before $s$, while no source edge from $e_\ell$ onward has occurred by $s$.
		Hence, $z = \sigma_\ell^-$.
		Since $R_\ell$ has a nonempty intersection with $I$ and contains $T_\ell \geq s'$, we have $a_\ell < s' < b_\ell$.
		So, $L_\ell$ is a prefix or infix language and contains $z$.
		In this case, choose $u_\ell = z$.
		In either case, choose $u_j = \epsilon$ for every other index.
		Since no occurrence lies in $(s,s')$, the preceding observation gives $u_j \in L_j$ for every unselected index.
		Therefore, $u_j \in L_j$ for every $j$ and $\destutter(u_1 \ldots u_m) = z = \pi_I(x_i^c)$.

		In both cases, the selected concatenation is nonempty and destutters to $\pi_I(x_i^c)$.
		The definition of $\gamma$ therefore gives $\pi_I(x_i^c) \in \gamma(x_i,I)$.
		Since $i$ and $I$ were arbitrary, every component $x_i^c$ satisfies every segmentwise condition in the definition of $\tr^+(S,\hb)$, so $w_c \in \tr^+(S,{\hb})$.
		Since $w_c$ was arbitrary, $\tr(S,{\hb}) \subseteq \tr^+(S,{\hb})$.
	\end{proof}

	\section{Monitoring Algorithms} 
	\label{sec:algorithm}
	
	Having presented the necessary abstractions, we now give symbolic monitoring algorithms in two different settings: the \emph{offline} setting where the entire distributed signal is available at once (\cref{sec:offline}) and the \emph{online} setting where the signal becomes available incrementally (\cref{sec:online}).
	In the offline setting, our algorithm works both for future STL and past STL.
	In the online setting, we consider only past STL.

	\subsection{Offline Algorithm}
	\label{sec:offline}

	In the offline setting, the entire distributed signal is available.
	For every subformula $\psi$ of a given STL formula $\varphi$ and every canonical segment $I \in G_S$, the monitor computes the finite symbolic output $\llbracket (S,{\hb}),I \models \psi \rrbracket$.
	It computes these sets from the function $\gamma$ without constructing $\tr^+(S,{\hb})$.
	After describing the algorithm, we prove that the symbolic output contains the value expressions corresponding to the satisfaction signals induced by every trace in $\tr^+(S,{\hb})$, i.e., the second inclusion in \cref{sec:semantics}:
	\[
	\{\pi_I(y_{\psi,w}) \st w \in \tr^+(S,{\hb})\}
	\subseteq
	\llbracket (S,{\hb}),I \models \psi \rrbracket.
	\]

	We first introduce the \emph{asynchronous product} of value expressions to capture interleavings within segments, then evaluate \emph{untimed} and \emph{timed} operators.
	Finally, we combine these steps into a \emph{symbolic evaluator} from which a sound monitoring verdict is extracted.
	We also discuss an efficient implementation of the monitoring algorithm using \emph{bit vectors}, efficient \emph{numerical abstractions} for handling real-valued signals, and a method to \emph{combine} our approach with exact monitoring.
	
	\paragraph*{Asynchronous Products.}
	Consider the value expressions $u_1 = 01$ and $u_2 = 10$ encoding the behaviors of two signals within a segment.
	Since behaviors within a segment are asynchronous, to capture their potential interleavings, we consider how the values in $u_1$ and $u_2$ can align.
	In particular, there are three potential alignments:
	(i) the rising edge of $u_1$ happens before the falling edge of $u_2$,
	(ii) the falling edge of $u_2$ happens before the rising edge of $u_1$, and
	(iii) they happen simultaneously.
	We respectively represent these with the tuples $(011, 110)$, $(001, 100)$, and $(01, 10)$ where the first component encodes $u_1$ and the second $u_2$.
	Formally, given $m \geq 1$ value expressions $u_1, \ldots, u_m$ (resp. sets $L_1, \ldots, L_m$ of value expressions), we define their \emph{asynchronous product} as follows, for $k \defeq 1 + \sum_{i=1}^{m} (|u_i| - 1)$:
	\begin{align*}
			\bigotimes_{i=1}^{m} u_i &\defeq
			\begin{aligned}[t]
				\{ \destutter(v_1,\ldots,v_m) \st v_i \in \stutter_k(u_i) \\
				\text{ for all } 1 \leq i \leq m \}
			\end{aligned}\\
			\bigotimes_{i=1}^{m} L_i &\defeq \bigcup_{(u_1,\ldots,u_m) \in L_1 \times \ldots \times L_m} \bigotimes_{i=1}^{m} u_i
	\end{align*}
	For $m = 2$, we write $u_1 \otimes u_2$ and $L_1 \otimes L_2$.
	Each element is a tuple of equal-length value expressions that describes one possible alignment, and the product contains the alignments induced by any actual tuple of signals.
	
	Asynchronous products of value expressions allow us to lift value expressions to satisfaction signals of formulas.
	
	\begin{example} \label{ex:asyncprod}
		Recall $(S, {\hb})$ and its $\gamma$ function given in \cref{fig:csve}.
		Here and below, for each signal $x_i$, we write $x_i$ also for the atomic proposition $x_i > 0$.
		To compute the value expressions encoding the satisfaction of $x_1 \land x_2$ in the segment $[3,4)$, we first compute the asynchronous product $\gamma(x_1, [3,4)) \otimes \gamma(x_2, [3,4))$, and then the bitwise conjunction of each pair in the set.
		For example, taking the expression $0 1 0$ for $x_1$ and $0 1$ for $x_2$, the product contains the pair $(010, 011)$.
		Its bitwise conjunction is $0 1 0$, encoding a potential behavior for the satisfaction of $x_1 \land x_2$.
	\end{example}

	\paragraph*{Untimed Operations.}
	As hinted in \cref{ex:asyncprod}, to compute the symbolic output, we apply bitwise operations on value expressions of Boolean signals and their asynchronous products to transform them into value expressions for the satisfaction signals of formulas.
	For example, to compute the segmentwise outputs for $\LTLeventually (x_1 \land x_2)$, we first compute for each segment in $G_S$ the set of value expressions for the satisfaction of $x_1 \land x_2$, and then from these compute those of $\LTLeventually (x_1 \land x_2)$.
	This compositional approach allows us to evaluate arbitrary STL formulas.

	\begin{figure*}[!t]
		\begin{align*}
			\llbracket (S, {\hb}), I \models \true \rrbracket
			&\defeq \{1\} \\
			\llbracket (S, {\hb}), I \models p \rrbracket
			% &\defeq \gamma(x_p, I) \\
			&\defeq \destutter\!\left(
				\left\{
					\operatorname{eval}_p(v_1,\ldots,v_n)
					\st
					(v_1,\ldots,v_n)
					\in
					\bigotimes_{i=1}^{n}\gamma(x_i,I)
				\right\}
			\right) \\
			\llbracket (S, {\hb}), I \models \lnot \varphi_1 \rrbracket
			&\defeq
			\{\BitNeg u \st u \in \llbracket (S, {\hb}), I \models \varphi_1 \rrbracket\} \\
			\llbracket (S, {\hb}), I \models \varphi_1 \land \varphi_2 \rrbracket
			&\defeq
			\destutter(
				\{
					u_1 \BitAnd u_2
					\st
					(u_1,u_2)
					\in
					\llbracket (S, {\hb}), I \models \varphi_1 \rrbracket
					\otimes
					\llbracket (S, {\hb}), I \models \varphi_2 \rrbracket
				\}
			) \\
			\llbracket (S, {\hb}), I \models \varphi_1 \until \varphi_2 \rrbracket
			&\defeq
			\destutter(
				\{
					u_1 \mathsf{U}^a u_2
					\st
					(u_1,u_2)
					\in
					\llbracket (S, {\hb}), I \models \varphi_1 \rrbracket
					\otimes
					\llbracket (S, {\hb}), I \models \varphi_2 \rrbracket, \\
			&\hspace{17em}
					a \in
					\first(
						\llbracket
							(S, {\hb}), I'
							\models
							\varphi_1 \until \varphi_2
						\rrbracket
					)
				\}
			) \\
			\llbracket (S, {\hb}), I \models \varphi_1 \since \varphi_2 \rrbracket
			&\defeq
			\destutter(
				\{
					u_1 \mathsf{S}^a u_2
					\st
					(u_1,u_2)
					\in
					\llbracket (S, {\hb}), I \models \varphi_1 \rrbracket
					\otimes
					\llbracket (S, {\hb}), I \models \varphi_2 \rrbracket, \\
			&\hspace{2em}
					a \in
					\left\{
						b \land c
						\st
						% \begin{array}{l}
							b \in
							\last\!\left(
								\llbracket
									(S, {\hb}), I''
									\models
									\varphi_1 \since \varphi_2
								\rrbracket
							\right),
							c \in
							\last\!\left(
								\llbracket
									(S, {\hb}), I''
									\models
									\varphi_1
								\rrbracket
							\right)
						% \end{array}
					\right\}
				\}
			)
		\end{align*}
		\caption{
			Inductive symbolic evaluation of untimed formulas.
			Here, $I'$ is the segment immediately following $I$ in $G_S$, and $I''$ is the segment immediately preceding $I$.
			To make the boundary cases uniform, we use the convention $\llbracket (S,{\hb}),J \models \varphi \rrbracket \defeq \{0\}$ for every formula $\varphi$ and every $J \notin G_S$.
			In $\until$, $a$ is the value at the start of $I'$.
			In $\since$, $a$ is the value carried over from $I''$'s open end.
			In either case, $a = 0$ gives the strong case and $a = 1$ the weak case.
			\label{fig:eval}
		}
	\end{figure*}

	First, we define bitwise operations on value expressions.
	Then, we use them to evaluate untimed formulas over sets of value expressions.
	Let $u$ and $v$ be Boolean value expressions of length $\ell$.
	We denote by $u \BitAnd v$ the bitwise-and operation, by $u \BitOr v$ the bitwise-or, and by $\BitNeg u$ the bitwise-negation.
	We also define the \emph{bitwise strong until} and \emph{bitwise strong since} operators:
	\begin{align*}
		u \mathsf{U}^0 v
		&\defeq
		\left(
			\max_{i \leq j \leq \ell}
			\left(
				\min
				\left(
					v[j],
					\min_{i \leq k < j}u[k]
				\right)
			\right)
		\right)_{1 \leq i \leq \ell} \\
		u \mathsf{S}^0 v
		&\defeq
		\left(
			\max_{1 \leq j \leq i}
			\left(
				\min
				\left(
					v[j],
					\min_{j < k \leq i}u[k]
				\right)
			\right)
		\right)_{1 \leq i \leq \ell}
	\end{align*}
	where maxima and minima are taken with respect to the order $0 < 1$, and $\min \emptyset \defeq 1$.
	In particular, for one-letter expressions we have $0 \mathsf{U}^0 1 = 1$ and $1 \mathsf{S}^0 1 = 1$.
	As usual, we derive \emph{bitwise eventually} as $\mathsf{E} u \defeq 1^\ell \mathsf{U}^0 u$, \emph{bitwise always} as $\mathsf{A}u \defeq \BitNeg(\mathsf{E} \BitNeg u)$, and \emph{bitwise weak until} as $u \mathsf{U}^1 v \defeq (u \mathsf{U}^0 v) \BitOr (\mathsf{A} u)$.
	The corresponding past-time operations are defined similarly:
	$\bar{\mathsf{E}} u \defeq 1^\ell \mathsf{S}^0 u$,
	$\bar{\mathsf{A}} u \defeq \BitNeg(\bar{\mathsf{E}} \BitNeg u)$, and
	$u \mathsf{S}^1 v \defeq (u \mathsf{S}^0 v) \BitOr (\bar{\mathsf{A}} u)$.
	The distinction between $\mathsf{U}^0$ and $\mathsf{U}^1$ (resp. $\mathsf{S}^0$ and $\mathsf{S}^1$) captures whether satisfaction may continue across the segment boundary, and it will be useful in the monitoring algorithms.
	These operations return a value expression of length $\ell$, not necessarily destuttered.
	For example, if $u = 010$, we have $\mathsf{E} u = 110$ and $\mathsf{A} u = 000$.

	Let $v_1, \ldots, v_n$ be equal-length real-valued value expressions of length $\ell$.
	For each atomic proposition $p \in \AP$, we define the Boolean value expression $\operatorname{eval}_p(v_1, \ldots, v_n)$ as  $1$ if $f_p(v_1[j], \ldots, v_n[j]) > 0$, and $0$ otherwise, for every $1 \leq j \leq \ell$.

	Let $(S,{\hb})$ be a distributed signal and let $\varphi_1,\varphi_2$ be two STL formulas.
	We define the symbolic evaluation of constants, atomic propositions, Boolean operators, and untimed temporal operators with respect to $(S,{\hb})$ and a segment $I \in G_S$ inductively as shown in \cref{fig:eval}.
	When $\varphi$ is a future (resp. past) formula and $I$ is the first (resp. last) segment in $G_S$, we write $\llbracket (S,{\hb}) \models \varphi \rrbracket$.

	For a given formula and segment, the evaluation above produces its symbolic output on the segment.
	Since the choices for the immediate subformulas and for adjacent segments are independent, the symbolic output can contain additional value expressions.

	\begin{example}
		Recall $(S,{\hb})$ and $\gamma$ from \cref{fig:csve}.
		To compute $\llbracket (S,{\hb}),[5,7) \models \LTLeventually(x_1 \land x_2) \rrbracket$, we first compute $\llbracket (S,{\hb}),[5,7) \models x_1 \land x_2 \rrbracket$ by taking the bitwise conjunction over the asynchronous product $\gamma(x_1,[5,7)) \otimes \gamma(x_2,[5,7))$ and destuttering.
		For example, since $10 \in \gamma(x_1,[5,7))$ and $10 \in \gamma(x_2,[5,7))$, the pair $(100,110)$ is in the product, whose conjunction gives us $10$ after destuttering.
		Repeating this for the rest, we obtain $\llbracket (S,{\hb}),[5,7) \models x_1 \land x_2 \rrbracket = \{0,10\}$.
		Note that the boundary bit inherited from the next segment is $a = 0$ since $\llbracket (S,{\hb}),[7,8) \models \LTLeventually (x_1 \land x_2) \rrbracket = \{0\}$.
		So, we compute $\llbracket (S,{\hb}),[5,7) \models \LTLeventually (x_1 \land x_2) \rrbracket$ by applying the bitwise eventually operator to each expression in $\llbracket (S,{\hb}),[5,7) \models x_1 \land x_2 \rrbracket$ and destuttering.
		The resulting set $\{0,10\}$ is the symbolic output of $\LTLeventually (x_1 \land x_2)$ on $[5,7)$.
	\end{example}

	\paragraph*{Timed Operations.}
	The untimed evaluation rule in \cref{fig:eval} is local because the interval $[0,\infty)$ admits a witness at any distance from the current time.
	All we need from outside the current segment is therefore a single bit summarizing whether the temporal condition can be propagated across the relevant segment boundary.
	For timed formulas such as $\varphi_1 \until_J \varphi_2$ with $J \neq [0,\infty)$, a time at which $\varphi_2$ holds is a witness only if its distance from the evaluation time lies in $J$; such a single bit cannot capture this since it records whether a witness exists but not at what distance.
	To evaluate at one time $t$, the evaluator may now need the satisfaction signals of the two subformulas on several canonical segments.
	Note, however, that these signals themselves are not available; for each subformula and each segment, the evaluator has only the symbolic output computed earlier, i.e., a finite set of value expressions describing the signal's possible behaviors on that segment.

	We describe the timed evaluation rule in two steps.
	For a fixed evaluation time $t$, we first combine these segmentwise outputs into a \emph{profile} of the left subformula on the \emph{horizon} (where it may need to hold before a witness) and that of the right subformula on the \emph{witness window} (where a witness may occur).
	We then track how these intervals, and with them the truth value of the formula, can change as $t$ ranges over the current segment, which we partition into finitely many \emph{placement regions}.

	Consider a distributed signal $(S,{\hb})$ with the temporal domain $[0,d)$, a segment $I \in G_S$, and an interval $J \subseteq \R_{\geq 0}$ with $J \neq [0,\infty)$.
	For every formula $\psi$ and segment $K \in G_S$, we write $\tau_{\psi,K} \defeq \llbracket (S,{\hb}), K \models \psi \rrbracket$ as shorthand.
	Let $F$ be the set of canonical-segment endpoints and let $F^{\circ} \defeq F \setminus \{0,d\}$.
	Given an interval $A \subseteq \R$, we write $l_A \defeq \inf A$ and $r_A \defeq \sup A$.

	We focus on the rule for $\until_J$ and define the past-time rule by the symmetric construction below.

	\subparagraph*{One Evaluation Time.}
	Consider an evaluation time $t \in I$.

	Let $W_t^{\rightarrow} \defeq (t \oplus J) \cap [0,d)$ be the set of times at which $\varphi_2$ may provide a witness, which we call the \emph{witness window}.
	A time $s \in W_t^{\rightarrow}$ is a witness if $\varphi_2$ holds at $s$ and $\varphi_1$ holds throughout $[t,s)$.
	To check this condition on $\varphi_1$, we use the \emph{horizon}
	\[
		H_t^{\rightarrow}
		\defeq
		\begin{cases}
			[t,t+\sup J] \cap [0,d) & \text{if } \sup J < \infty,\\
			[t,d) & \text{if } \sup J = \infty.
		\end{cases}
	\]

	Since either interval may cross several canonical segments, we define an operation that restricts a segment's output to the part inside the interval.
	For a language $L$, we write $\pfx_+(L) \defeq \pfx(L) \setminus \{\epsilon\}$, $\sfx_+(L) \defeq \sfx(L) \setminus \{\epsilon\}$, and $\infx_+(L) \defeq \infx(L) \setminus \{\epsilon\}$.
	Let $A \subseteq \R$ be an interval, let $K = [l_K,r_K) \in G_S$, and let $M \defeq A \cap K$ be the part of $K$ that lies inside $A$.
	With $L \defeq \tau_{\psi,K}$, let us define
	\[
		\mathsf{res}(\psi,K,M)
		\defeq
		\left\{
		\begin{array}{@{}l@{\;}l@{}}
			L
			& \text{if } M = K,\\
			\first(L)
			& \text{if } M = \{l_K\},\\
			\pfx_+(L)
			& \begin{aligned}[t]
				&\text{if } l_K < r_M < r_K\\
				&\text{and } l_K \in M,
			\end{aligned}\\
			\sfx_+(L)
			& \begin{aligned}[t]
				&\text{if } r_M = r_K\\
				&\text{and } l_K \notin M,
			\end{aligned}\\
			\infx_+(L)
			& \text{otherwise.}
		\end{array}
		\right.
	\]
	These cases cover all the ways in which an interval $M$ can lie inside the segment $K$.
	If a signal has a value expression in $L$ on segment $K$, then its value expression on $M$ belongs to $\mathsf{res}(\psi,K,M)$.
	Conceivably, the operation can also result in additional value expressions because a value expression records the order of changes but not when they occur.

	Let $K_1 < \ldots < K_q$ be the canonical segments that intersect $A$.
	The \emph{profile} of $\psi$ on $A$ is
	\[
	\begin{aligned}
		\mathsf{prof}(\psi,A)
		\defeq
		\destutter \bigl(&
			\mathsf{res}(\psi,K_1,A \cap K_1)
			\cdot \ldots \cdot {} \\[-1mm]
			&
			\mathsf{res}(\psi,K_q,A \cap K_q)
		\bigr).
	\end{aligned}
	\]
	To obtain a value expression in $\mathsf{prof}(\psi,A)$, we choose one expression from the corresponding $\mathsf{res}$ set for each segment that intersects $A$, concatenate the choices in temporal order, and destutter the result.
	The profile soundly overapproximates the value expressions of the satisfaction signal of $\psi$ on $A$.
	In particular, for every trace $w$, if $\pi_K(y_{\psi,w}) \in \tau_{\psi,K}$ for every segment $K$ with $K \cap A \neq \emptyset$, then $\pi_A(y_{\psi,w}) \in \mathsf{prof}(\psi,A)$.

	Note that we build the profile of $\varphi_1$ on the horizon $H_t^{\rightarrow}$ since it contains all the time points at which $\varphi_1$ may need to hold before a witness, and that of $\varphi_2$ on the witness window $W_t^{\rightarrow}$ since it contains exactly the admissible witness times.

	\begin{example}\label{ex:timed-profile}
		Recall the distributed signal in \cref{fig:csve}.
		Let $\varphi_1 \defeq \lnot x_1$ and $\varphi_2 \defeq x_2$.
		For the first two segments $K_0 \defeq [0,1)$ and $K_1 \defeq [1,3)$, \cref{fig:csve} gives
		\[
			\begin{aligned}
				\tau_{\varphi_1,K_0} &= \{1,10\}, &
				\tau_{\varphi_1,K_1} &= \{0,1,10\},\\
				\tau_{\varphi_2,K_0} &= \{0\}, &
				\tau_{\varphi_2,K_1} &= \{0,01\}.
			\end{aligned}
		\]
		Let $I = K_0$, $J = [0,1)$, and $t = \tfrac{1}{2}$.
		Then, $W_t^{\rightarrow} = [\tfrac{1}{2},\tfrac{3}{2})$ and $H_t^{\rightarrow} = [\tfrac{1}{2},\tfrac{3}{2}]$.
		The horizon intersects $K_0$ in a suffix and $K_1$ in a prefix.
		The witness window also gives a suffix of $K_0$ and a prefix of $K_1$.
		Therefore,
		\[
			\begin{gathered}
				\mathsf{prof}(\varphi_1,H_t^{\rightarrow})\\
				= \destutter \bigl(
				\sfx_+(\{1,10\}) \cdot \pfx_+(\{0,1,10\})
				\bigr)\\
				= \{0,01,010,1,10,101,1010\},
			\end{gathered}
		\]
		and
		\[
			\begin{gathered}
				\mathsf{prof}(\varphi_2,W_t^{\rightarrow})\\
				= \destutter \bigl(
				\sfx_+(\{0\}) \cdot \pfx_+(\{0,01\})
				\bigr)\\
				= \{0,01\}.
			\end{gathered}
		\]
	\end{example}

	\subparagraph*{Finitely Many Placements.}
	The segment $I$ contains infinitely many evaluation times, but only finitely many placements of the horizon and witness window relative to the canonical segments.
	In particular, as $t$ moves through $I$, how the horizon and the witness window intersect with the canonical segments can change only when one of their boundaries reaches an endpoint of a canonical segment, which can happen finitely many times since there are finitely many boundaries and endpoints.
	The witness-window boundary offsets are
	\[
		C_J
		\defeq
		\begin{cases}
			\{\inf J,\sup J\}
			& \text{if } \sup J < \infty,\\
			\{\inf J\}
			& \text{if } \sup J = \infty.
		\end{cases}
	\]
	The relevant offsets for the horizon and the witness window are
	\[
		M_J \defeq \{0\} \cup C_J.
	\]

	A time $t \in I$ is \emph{critical} if $t+c \in F$ for some $c \in M_J$.
	Since $F$ and $M_J$ are finite, there are only finitely many critical times.

	We keep each critical time as a singleton region and use the maximal open intervals between them as the remaining regions.
	Let $T_1 < \ldots < T_k$ be the resulting \emph{placement regions}.
	We include the singleton regions, because at those times an intersection with a canonical segment can use a different case of $\mathsf{res}$ than immediately before or after it.

	For every placement region $T_j$ and all $t,t' \in T_j$, the horizons $H_t^{\rightarrow}$ and $H_{t'}^{\rightarrow}$ intersect the canonical segments in the same way, and so do the witness windows $W_t^{\rightarrow}$ and $W_{t'}^{\rightarrow}$.
	Hence, the profiles used below are the same for all times in $T_j$.
	The two horizons can also be mapped onto each other in temporal order so that $t$ is mapped to $t'$ and $W_t^{\rightarrow}$ is mapped to $W_{t'}^{\rightarrow}$.
	This mapping preserves the value expressions and the timed-until result, so the same truth values are possible at every time in $T_j$.
	We can therefore compute them using one representative $t_j \in T_j$, although a fixed trace may still change its truth value within $T_j$.

	Let $t_j \in T_j$.
	We define
	\[
	\begin{aligned}
		H_j &\defeq H_{t_j}^{\rightarrow}, &
		W_j &\defeq W_{t_j}^{\rightarrow},\\
		P_j &\defeq \mathsf{prof}(\varphi_1,H_j).
	\end{aligned}
	\]
	If $W_j = \emptyset$, let $B_j \defeq \{0\}$.
	Otherwise, let $Q_j \defeq \mathsf{prof}(\varphi_2,W_j)$, and let $B_j$ be the set of bits $b \in \B$ for which there are signals $z_1 : H_j \to \B$ and $z_2 : W_j \to \B$ such that
	\begin{enumerate}
		\item $\pi_{H_j}(z_1) \in P_j$,
		\item $\pi_{W_j}(z_2) \in Q_j$, and
		\item $b = 1$ iff there exists $s \in W_j$ with $z_2(s) = 1$ and $z_1(r) = 1$ for all $r \in [t_j,s)$.
	\end{enumerate}
	When $W_j$ is nonempty, the signals $z_1$ and $z_2$ range over all signals described by the profiles $P_j$ and $Q_j$.
	Since the profiles are overapproximations and the two signals are chosen independently, $B_j$ may then contain a bit that no trace produces.
	In either case, if the segmentwise outputs contain the value expressions induced by a trace $w$, then $y_{\varphi_1 \until_J \varphi_2,w}(t) \in B_j$ for every $t \in T_j$.
	
	Note that $B_j$ is computable even though $z_1$ and $z_2$ range over infinitely many signals.
	For each pair of value expressions in $P_j$ and $Q_j$, only the order of their changes relative to each other and to the endpoints of $H_j$ and $W_j$ matters.
	There are only finitely many such orders, so we check them and collect the resulting bits in $B_j$.

	\begin{figure*}[!t]
		\begin{align*}
			\llbracket (S,{\hb}),I \models \varphi_1 \until_J \varphi_2 \rrbracket
			&\defeq \destutter \bigl( \Alt_{N_1}(B_1) \cdot \ldots \cdot \Alt_{N_k}(B_k) \bigr), \\
			\llbracket (S,{\hb}),I \models \varphi_1 \since_J \varphi_2 \rrbracket
			&\defeq \destutter \bigl( \Alt_{N'_1}(B'_1) \cdot \ldots \cdot \Alt_{N'_\ell}(B'_\ell) \bigr).
		\end{align*}
		\caption{
			Symbolic evaluation of timed until and timed since over the current segment $I$ and timing interval $J$.
			For $\varphi_1 \until_J \varphi_2$, the regions $T_1 < \ldots < T_k$ partition $I$ according to the possible placements of the future witness window $W$ and horizon $H$.
			The set $B_j$ contains the possible truth values at a time in $T_j$, and $N_j$ bounds the length of the value expression on that region.
			The primed objects $T'_1 < \ldots < T'_\ell$, $B'_j$, and $N'_j$ have the same roles for $\varphi_1 \since_J \varphi_2$ and its past witness window and horizon.
			The operation $\Alt_N(B)$ forms all alternating value expressions over $B$ of length at most $N$.
			The region-wise expressions are concatenated in temporal order, and $\destutter$ removes repeated values at their boundaries.
			\label{fig:timedEval}
		}
	\end{figure*}

	\begin{example}\label{ex:timed-bits}
		Let us continue \cref{ex:timed-profile} with the formula $\varphi_1 \until_{[0,1)} \varphi_2 = \lnot x_1 \until_{[0,1)} x_2$.
		The only critical time in $I = [0,1)$ is $0$, so the placement regions are $T_1 = \{0\}$ and $T_2 = (0,1)$.

		For $T_1$, take $t_1 = 0$.
		Then, $H_1 = [0,1]$ and $W_1 = [0,1)$, and the two profiles are $P_1 = \{1,10,101\}$ and $Q_1 = \{0\}$.
		Since $\varphi_2$ is false throughout the witness window, we get $B_1 = \{0\}$.

		For $T_2$, take $t_2 = \tfrac{1}{2}$.
		Recall that $W_2 = [\tfrac{1}{2},\tfrac{3}{2})$.
		By \cref{ex:timed-profile}, we have $P_2 = \{0,01,010,1,10,101,1010\}$ and $Q_2 = \{0,01\}$.
		The expression $0 \in Q_2$ gives no witness.
		The expression $01 \in Q_2$ means that $\varphi_2$ becomes true in $W_2$.
		It gives a witness if a signal described by $P_2$ remains true until that change, and no witness otherwise.
		Therefore, $B_2 = \{0,1\}$.
	\end{example}

	\subparagraph*{Value Expressions on a Placement Region.}
	The set $B_j$ contains every bit that can occur at an individual time in $T_j$, but it does not describe how the value can change across the region.
	We now derive a bound on the number of changes from the segmentwise outputs of the two subformulas.

	For a formula $\psi$ and an interval $A \subseteq \R$, we define
	\[
		E_\psi(A)
		\defeq
		|F^{\circ} \cap A|
		+
		\sum_{\substack{K \in G_S \\ K \cap A \neq \emptyset}}
		\left( \max_{u \in \tau_{\psi,K}}|u| - 1 \right).
	\]
	A value expression of length $m$ records $m-1$ changes.
	The sum therefore bounds the changes inside the canonical segments that have a nonempty intersection with $A$, while $|F^{\circ} \cap A|$ allows one additional change at each internal segment endpoint.
	Hence, $E_\psi(A)$ bounds the number of changes on $A \cap [0,d)$ of every satisfaction signal whose value expression on each crossed segment $K$ belongs to $\tau_{\psi,K}$.

	If $T_j$ is a singleton, let $N_j \defeq 1$.
	Otherwise, let
	\[
		\begin{aligned}
			N_j \defeq 1 + 2 \biggl(
			&\sum_{c \in M_J} E_{\varphi_1}(T_j \oplus c) \\
			&+ \sum_{c \in C_J} E_{\varphi_2}(T_j \oplus c)
			\biggr).
		\end{aligned}
	\]
	For $\varphi_1$, all moving boundaries with offsets in $M_J$ matter, while, for $\varphi_2$, only the witness-window boundaries with offsets in $C_J$ matter.
	As $t$ moves through $T_j$, the two sums bound the number $D$ of times at which one of these boundaries reaches a change of the corresponding satisfaction signal.
	Between these times, the timed-until value is constant.
	The $D$ times divide $T_j$ into at most $D+1$ open intervals and may each add one value.
	The resulting value expression therefore has at most $2D+1$ letters, which gives the bound $N_j$.

	For $N \geq 1$ and $B \subseteq \B$, we define
	\[
		\Alt_N(B) \defeq \destutter \left( \bigcup_{\ell=1}^{N} B^\ell \right)
	\]
	as the set of Boolean value expressions over $B$ with at most $N$ letters.
	It follows that $\Alt_{N_j}(B_j)$ contains every value expression of the timed-until satisfaction signal on $T_j$.

	The symbolic outputs for timed until and timed since are defined in
	\cref{fig:timedEval}.
	The factors cover the placement regions in temporal order.
	Their concatenation combines the region-wise value expressions, and destuttering removes repeated values at the boundaries between consecutive regions.

	\begin{example}\label{ex:timed-eval}
		Let us continue \cref{ex:timed-bits} and complete the evaluation of $\lnot x_1 \until_{[0,1)} x_2$ over $I = [0,1)$.
		Recall that the placement regions are $T_1 = \{0\}$ and $T_2 = (0,1)$, with $B_1 = \{0\}$ and $B_2 = \{0,1\}$.

		Since $T_1$ is a singleton, we have $N_1 = 1$.
		For $T_2$, we have $M_J = C_J = \{0,1\}$.
		The left subformula $\varphi_1 = \lnot x_1$ can change once on $T_2$ and once on $T_2 \oplus 1$, while the right subformula $\varphi_2 = x_2$ is constant on $T_2$ and can change once on $T_2 \oplus 1$.
		Thus, $E_{\varphi_1}(T_2) = E_{\varphi_1}(T_2 \oplus 1) = 1$, as well as $E_{\varphi_2}(T_2) = 0$ and $E_{\varphi_2}(T_2 \oplus 1) = 1$.
		It follows that
		\[
			\begin{aligned}
				N_2
				&= 1 + 2 \Biggl(
					\sum_{c \in \{0,1\}} E_{\varphi_1}(T_2 \oplus c) \\
				&\hspace{5.5em}
					+ \sum_{c \in \{0,1\}} E_{\varphi_2}(T_2 \oplus c)
				\Biggr)
				= 7.
			\end{aligned}
		\]
		We therefore obtain $\Alt_{N_1}(B_1) = \{0\}$ and $\Alt_{N_2}(B_2) = \Alt_7(\{0,1\})$.
		Concatenating these sets and destuttering gives
		\[
			\begin{aligned}
			&\llbracket (S,{\hb}),[0,1)
			\models \lnot x_1 \until_{[0,1)} x_2 \rrbracket\\
			&\quad = \{0,01,010,0101,01010,010101,{}\\[-1mm]
			&\hspace{30mm}0101010,01010101\}.
			\end{aligned}
		\]
		The initial $0$ is forced by $T_1$, since $x_2$ is false throughout the witness window $[0,1)$ at time $0$.

		Notice that only $0$ and $01$ are realizable.
		Let $s_1$ and $s_2$ be the reference times at which $x_1$ and $x_2$ first change from $0$ to $1$.
		Since the uncertainty region of $x_2$'s rising edge is $(1,5)$, we have $s_2 \in (1,5)$;so, if $s_2 \notin (1,2)$, then $s_2 \geq 2$, and no window $[t,t+1)$ with $t \in [0,1)$ contains $s_2$.
		If $s_2 \in (1,2)$ and $s_1 < s_2$, the left subformula becomes false before $x_2$ becomes true.
		In both cases, the formula is always false on $I$, and its value expression is $0$.
		If $s_2 \in (1,2)$ and $s_1 \geq s_2$, the formula is false for $t \leq s_2-1$ and true for $t > s_2-1$, so its value expression is $01$.
		Once the formula becomes true on $I$, it cannot become false again.
		Hence, all remaining value expressions in the symbolic output are spurious.
	\end{example}

	\begin{figure*}[!t]
		\centering
		\begin{tikzpicture}[
			x=.945cm,
			y=1cm,
			font=\footnotesize,
			panel/.style={
				draw=gray!45,
				fill=gray!3,
				rounded corners=2pt,
				line width=.5pt
			},
			segment/.style={
				draw=gray!45,
				fill=gray!7,
				line width=.7pt
			},
			output/.style={
				draw=gray!45,
				fill=white,
				line width=.5pt
			},
			horizon/.style={black,line width=1.1pt},
			window/.style={admAccentBlue,line width=1.6pt}
		]
			\node[anchor=west,font=\small\bfseries] at (0,.25)
				{(a) Inputs and partition of $I$};

			\node[anchor=east] at (1.7,-.55) {canonical segment};
			\draw[segment] (2,-.75) rectangle (4.4,-.35);
			\draw[segment] (4.4,-.75) rectangle (9.2,-.35);
			\node at (3.2,-.55) {$K_0=[0,1)$};
			\node at (6.8,-.55) {$K_1=[1,3)$};
			\foreach \x/\lab in {2/0,4.4/1,9.2/3}
				\node[above=1pt] at (\x,-.35) {$\lab$};

			\node[anchor=east] at (1.7,-.95) {$\varphi_1=\lnot p_1$};
			\draw[output] (2,-1.15) rectangle (4.4,-.75);
			\draw[output] (4.4,-1.15) rectangle (9.2,-.75);
			\node at (3.2,-.95) {$\tau_{\varphi_1,K_0}=\{1,10\}$};
			\node at (6.8,-.95) {$\tau_{\varphi_1,K_1}=\{0,1,10\}$};

			\node[anchor=east] at (1.7,-1.35) {$\varphi_2=p_2$};
			\draw[output] (2,-1.55) rectangle (4.4,-1.15);
			\draw[output] (4.4,-1.55) rectangle (9.2,-1.15);
			\node at (3.2,-1.35) {$\tau_{\varphi_2,K_0}=\{0\}$};
			\node at (6.8,-1.35) {$\tau_{\varphi_2,K_1}=\{0,01\}$};

			\node[anchor=west,font=\bfseries] at (11.15,-.55)
				{$I=K_0=T_1\cup T_2$};
			\fill (10,-1) circle (1.6pt);
			\node[anchor=west] at (12.75,-1)
				{$T_1=\{0\}$ \enspace(critical)};
			\draw[line width=1pt] (10,-1.45) -- (12.4,-1.45);
			\draw[fill=white,line width=.65pt] (10,-1.45) circle (1.7pt);
			\draw[fill=white,line width=.65pt] (12.4,-1.45) circle (1.7pt);
			\node[anchor=west] at (12.75,-1.45) {$T_2=(0,1)$};

			\node[anchor=west,font=\small\bfseries] at (0,-1.95)
				{(b) Evaluation at representative times};

			\draw[panel] (0,-5.45) rectangle (7.65,-2.25);
			\node[anchor=west,font=\bfseries] at (2,-2.58)
				{$T_1=\{0\}\qquad t_1=0$};

			\draw[gray!45] (.9,-3.12) -- (7,-3.12);
			\draw[gray!45] (1,-3.23) -- (1,-3.01);
			\draw[gray!45,dashed] (4,-3.98) -- (4,-2.92);
			\node[above=1pt] at (1,-3.12) {$0$};
			\node[above=1pt] at (4,-3.12) {$1$};
			\draw[horizon] (1,-3.45) -- (4,-3.45);
			\fill (1,-3.45) circle (1.4pt);
			\fill (4,-3.45) circle (1.4pt);
			\node[anchor=west] at (4,-3.45) {$H_1=[0,1]$};
			\draw[window] (1,-3.88) -- (4,-3.88);
			\fill[admAccentBlue] (1,-3.88) circle (1.4pt);
			\draw[admAccentBlue,fill=white,line width=.65pt]
				(4,-3.88) circle (1.6pt);
			\node[anchor=west,admAccentBlue] at (4,-3.88)
				{$W_1=[0,1)$};

			\node[anchor=north west,align=left,font=\footnotesize,
				text width=6.95cm] at (.35,-4.25)
				{$H_1 \cap K_0 = [0,1)$ \qquad $H_1 \cap K_1 = \{1\}$\\[.6mm]
				$P_1 = \mathsf{prof}(\varphi_1,H_1) = \{1,10,101\}$\\
				$Q_1 = \mathsf{prof}(\varphi_2,W_1) = \{0\}$\\};

			\draw[panel] (8.05,-5.45) rectangle (15.7,-2.25);
			\node[anchor=west,font=\bfseries] at (9.7,-2.58)
				{$T_2=(0,1) \qquad t_2=1/2$};

			\draw[gray!45] (8.95,-3.12) -- (15.1,-3.12);
			\draw[gray!45] (8.95,-3.23) -- (8.95,-3.01);
			\draw[gray!45,dashed] (11.95,-3.98) -- (11.95,-2.92);
			\node[above=1pt] at (8.95,-3.12) {$0$};
			\node[above=1pt] at (11.95,-3.12) {$1$};
			\draw[horizon] (10.45,-3.45) -- (13.45,-3.45);
			\fill (10.45,-3.45) circle (1.4pt);
			\fill (13.45,-3.45) circle (1.4pt);
			\node[anchor=west] at (13.45,-3.45)
				{$H_2=[1/2,3/2]$};
			\draw[window] (10.45,-3.88) -- (13.45,-3.88);
			\fill[admAccentBlue] (10.45,-3.88) circle (1.4pt);
			\draw[admAccentBlue,fill=white,line width=.65pt]
				(13.45,-3.88) circle (1.6pt);
			\node[anchor=west,admAccentBlue] at (13.45,-3.88)
				{$W_2=[1/2,3/2)$};

			\node[anchor=north west,align=left,font=\footnotesize,
				text width=6.95cm] at (8.4,-4.25)
				{$H_2 \cap K_0 = [1/2,1)$ \qquad $H_2 \cap K_1 = [1,3/2]$\\[.6mm]
				$P_2 = \mathsf{prof}(\varphi_1,H_2) = \{0,01,010,1,10,101,1010\}$\\
				$Q_2 = \mathsf{prof}(\varphi_2,W_2) = \{0,01\}$\\};

			\node[anchor=west,font=\small\bfseries] at (0,-5.92)
				{(c) Result on $I$};
			\node[anchor=north,font=\footnotesize] at (7.85,-6.18) {
				\renewcommand{\arraystretch}{1.25}
				\begin{tabular}{c@{\hspace{14mm}}c@{\hspace{14mm}}c@{\hspace{14mm}}l}
					$T_j$ & $B_j$ & $N_j$ & $\Alt_{N_j}(B_j)$ \\
					\hline
					$T_1$ & $\{0\}$ & $1$ & $\{0\}$ \\
					$T_2$ & $\{0,1\}$ & $7$ & $\Alt_7(\{0,1\})$
				\end{tabular}
			};

		\draw[draw=gray!45,fill=gray!3,rounded corners=2pt,line width=.6pt]
			(.25,-8.9) rectangle (15.45,-7.7);
		\node at (7.85,-8.3)
				{$\displaystyle
				\begin{aligned}
				\llbracket (S,{\hb}),I \models \lnot p_1 \until_{[0,1)} p_2 \rrbracket
				&=\destutter\bigl(\Alt_{N_1}(B_1)\cdot\Alt_{N_2}(B_2)\bigr)\\[-.5mm]
				&=\{0,01,010,0101,01010,010101,0101010,01010101\}
				\end{aligned}$};
		\end{tikzpicture}

		\caption{Symbolic evaluation of $\lnot p_1 \until_{[0,1)} p_2$ on $I=[0,1)$ from \cref{ex:timed-profile,ex:timed-bits,ex:timed-eval}.
		Panel (a) lists the previously computed sets $\tau_{\varphi_i,K}$ and the partition $I = T_1 \cup T_2$.
		Panel (b) shows $P_j$ on $H_j$ and $Q_j$ on $W_j$ at $t_1$ and $t_2$.
		Panel (c) computes $\Alt_{N_j}(B_j)$ and the final symbolic output.
		The value expressions $0$ and $01$ are realizable; the others are spurious.}
		\label{fig:timedOverview}
	\end{figure*}

	\subparagraph*{Past Time.}
	For $\varphi_1 \since_J \varphi_2$ and an evaluation time $t \in I$, let $W_t^{\leftarrow} \defeq (t \ominus J) \cap [0,d)$ be the past witness window, and let
	\[
		H_t^{\leftarrow}
		\defeq
		\begin{cases}
			[t-\sup J,t] \cap [0,d) & \text{if } \sup J < \infty,\\
			[0,t] & \text{if } \sup J = \infty
		\end{cases}
	\]
	be the past horizon.
	A time $s \in W_t^{\leftarrow}$ is a witness if $\varphi_2$ holds at $s$ and $\varphi_1$ holds throughout $(s,t]$.

	A time $t \in I$ is \emph{past-critical} if $t-c \in F$ for some $c \in M_J$.
	As above, we keep each past-critical time as a singleton region and use the maximal open intervals between them as the remaining regions.
	Let $T'_1 < \ldots < T'_\ell$ be the resulting past placement regions.
	For each $1 \leq j \leq \ell$, choose $t'_j \in T'_j$ and let $H'_j \defeq H_{t'_j}^{\leftarrow}$, $W'_j \defeq W_{t'_j}^{\leftarrow}$, and $P'_j \defeq \mathsf{prof}(\varphi_1,H'_j)$.

	If $W'_j = \emptyset$, let $B'_j \defeq \{0\}$.
	Otherwise, let $Q'_j \defeq \mathsf{prof}(\varphi_2,W'_j)$, and let $B'_j$ be the set of bits $b \in \B$ for which there are signals $z_1 : H'_j \to \B$ and $z_2 : W'_j \to \B$ such that
	\begin{enumerate}
		\item $\pi_{H'_j}(z_1) \in P'_j$,
		\item $\pi_{W'_j}(z_2) \in Q'_j$, and
		\item $b = 1$ iff there exists $s \in W'_j$ with $z_2(s) = 1$ and $z_1(r) = 1$ for every $r \in (s,t'_j]$.
	\end{enumerate}
	If $T'_j$ is a singleton, let $N'_j \defeq 1$.
	Otherwise, let
	\[
		\begin{aligned}
			N'_j \defeq 1 + 2 \biggl(
			&\sum_{c \in M_J} E_{\varphi_1}(T'_j \ominus c) \\
			&+ \sum_{c \in C_J} E_{\varphi_2}(T'_j \ominus c)
			\biggr).
		\end{aligned}
	\]
	The second equation in \cref{fig:timedEval} then gives the symbolic output for $\varphi_1 \since_J \varphi_2$.

	\paragraph*{Soundness of the Symbolic Evaluation.}
	As we have demonstrated in the examples above, the symbolic evaluation may include spurious outputs.
	We prove below that it nevertheless gives only sound verdicts.

	\begin{theorem} \label{cl:algo}
		Let $(S,{\hb})$ be a distributed signal and let $\varphi$ be an STL formula.
		Let $f \defeq \first$ if $\varphi$ is a future formula, and let $f \defeq \last$ if $\varphi$ is a past formula.
		Then,
		\begin{align*}
			f\bigl(\llbracket (S,{\hb}) \models \varphi \rrbracket\bigr)=\{1\}
			&\implies [(S,{\hb}) \models \varphi]=\top,\\
			f\bigl(\llbracket (S,{\hb}) \models \varphi \rrbracket\bigr)=\{0\}
			&\implies [(S,{\hb}) \models \varphi]=\bot.
		\end{align*}
	\end{theorem}

	The proof uses the second inclusion in \cref{sec:semantics}, which we show below.

	\begin{lemma} \label{cl:evalsound}
		For every distributed signal $(S,{\hb})$, STL formula $\psi$, and canonical segment $I \in G_S$,
		\[
			\{\pi_I(y_{\psi,w}) \st w \in \tr^+(S,{\hb})\}
			\subseteq
			\llbracket (S,{\hb}),I \models \psi \rrbracket.
		\]
	\end{lemma}
	\begin{proof}[\normalsize Proof.]
		\normalsize
		Let $w = (x'_1, \ldots, x'_n) \in \tr^+(S,{\hb})$.
		We prove the claim for all canonical segments by induction on $\psi$.
		For untimed until and since, we also use an induction over the segments.

		\emph{Let $\psi = \true$.}
		Then, $\pi_I(y_{\psi,w}) = 1$.

		\emph{Let $\psi = p$.}
		By the definition of $\tr^+(S,{\hb})$, we have $\pi_I(x'_i) \in \gamma(x_i,I)$ for every component $i$.
		Consider the partition $I$ into the maximal pieces on which all component signals are constant.
		For each $i$, let $v_i$ list the values of $x'_i$ on these pieces.
		By the definition of the asynchronous product, we have $(v_1,\ldots,v_n) \in \bigotimes_{i=1}^n \gamma(x_i,I)$.
		On each piece, $\operatorname{eval}_p(v_1,\ldots,v_n)$ is exactly the value of $y_{p,w}$.
		Therefore, $\pi_I(y_{p,w}) = \destutter \left( \operatorname{eval}_p(v_1,\ldots,v_n) \right)$, which is included in the symbolic output by the rule for $p$.

		\emph{Let $\psi = \lnot \psi_1$.}
		This case follows directly from the induction hypothesis and bitwise negation.

		\emph{Let $\psi = \psi_1 \land \psi_2$.}
		Consider the partition $I$ into the maximal pieces on which both satisfaction signals are constant, and let $v_1$ and $v_2$ list their values on these pieces.
		By the induction hypothesis and the definition of the asynchronous product, we have $(v_1,v_2) \in \llbracket (S,{\hb}),I \models \psi_1 \rrbracket \otimes \llbracket (S,{\hb}),I \models \psi_2 \rrbracket$.
		Since conjunction is evaluated pointwise, $\pi_I(y_{\psi,w}) = \destutter(v_1 \BitAnd v_2)$, which is included in the symbolic output by the rule for conjunction.

		\emph{Let $\psi = \psi_1 \until \psi_2$.}
		Let $I_1 < \ldots < I_m$ be the canonical segments.
		We proceed from $I_m$ to $I_1$.
		On $I_r$, let $a = 0$ if $r = m$; otherwise, let $a$ be the value of $y_{\psi,w}$ at the start of $I_{r+1}$.
		For $r = m$, we have $a = 0$ by definition.
		For $r < m$, the induction over segments gives $a \in \first\!\left(\llbracket (S,{\hb}),I_{r+1} \models \psi \rrbracket \right)$.

		By the structural induction hypothesis, the value expressions of $y_{\psi_1,w}$ and $y_{\psi_2,w}$ on $I_r$ belong to their respective symbolic outputs.
		Unlike in the pointwise cases above, their common refinement need not directly yield the value expression of until: a witness inside an open interval may also require $\psi_1$ on an earlier part of that piece.
		The asynchronous product contains an alignment $(v_1,v_2)$ that accounts for this distinction and satisfies $\pi_{I_r}(y_{\psi,w}) = \destutter(v_1\mathsf{U}^a v_2)$.
		Thus, the rule for until includes $\pi_{I_r}(y_{\psi,w})$.

		\emph{Let $\psi = \psi_1 \since \psi_2$.}
		We proceed from $I_1$ to $I_m$.
		Write $I_r = [s,t)$ and define $a_s = 1$ iff there exists $q \in [0,s)$ such that $y_{\psi_2,w}(q) = 1$ and $y_{\psi_1,w}(v) = 1$ for all $v \in (q,s)$.
		If $r = 1$, then $a_s = 0$, and both predecessor outputs are $\{0\}$ by definition.
		Otherwise, let $b \defeq \last\!\left(\pi_{I_{r-1}}(y_{\psi,w})\right)$ and $c \defeq \last\!\left(\pi_{I_{r-1}}(y_{\psi_1,w})\right)$.
		Then, $a_s = b \land c$.
		Indeed, if $a_s = 1$, then both signals equal $1$ on a final part of $I_{r-1}$.
		Conversely, if $b = c = 1$, choose $z < s$ sufficiently close to $s$ that both signals equal $1$ on $[z,s)$.
		A witness for $\psi$ at $z$, together with $y_{\psi_1,w} = 1$ on $(z,s)$, satisfies the definition of $a_s$.

		The induction over segments and the structural induction hypothesis give $b \in	\last\!\left( \llbracket (S,{\hb}),I_{r-1} \models \psi \rrbracket\right)$ and $c \in \last\!\left(\llbracket (S,{\hb}),I_{r-1} \models \psi_1 \rrbracket\right)$.
		Hence, the rule for since can choose $a_s$.
		By the structural induction hypothesis and the alignment argument above, the asynchronous product contains $(v_1,v_2)$ such that $\pi_{I_r}(y_{\psi,w}) = \destutter(v_1\mathsf{S}^{a_s}v_2)$.
		Thus, the rule for since includes $\pi_{I_r}(y_{\psi,w})$.

		\emph{Let $\psi = \psi_1 \until_J \psi_2$ with $J \neq [0,\infty)$.}
		Let $y_i \defeq y_{\psi_i,w}$ for $i \in \{1,2\}$, let $y \defeq y_{\psi,w}$, let $T_1 < \ldots < T_k$ be the placement regions on $I$, and let $j \in \{1,\ldots,k\}$.

		We first show that $y(t) \in B_j$ for every $t \in T_j$.
		Let $t \in T_j$.
		If $W_t^{\rightarrow} = \emptyset$, then $W_j = \emptyset$, so $y(t) = 0 \in B_j$.

		Otherwise, for every canonical segment $K$ that intersects $H_t^{\rightarrow}$, the induction hypothesis gives
		$\pi_K(y_1) \in \llbracket (S,{\hb}),K \models \psi_1 \rrbracket$.
		Similarly, for every canonical segment $K$ that intersects $W_t^{\rightarrow}$, it gives $\pi_K(y_2) \in \llbracket (S,{\hb}),K \models \psi_2 \rrbracket$.
		Therefore, $\pi_{H_t^{\rightarrow}}(y_1) \in \mathsf{prof}(\psi_1,H_t^{\rightarrow})$ and $\pi_{W_t^{\rightarrow}}(y_2) \in \mathsf{prof}(\psi_2,W_t^{\rightarrow})$.

		Since $t$ and $t_j$ lie in the same placement region, there is an order-preserving map from $H_t^{\rightarrow}$ to $H_j$ that maps $t$ to $t_j$ and $W_t^{\rightarrow}$ to $W_j$.
		Use this map to copy $y_1$ on $H_t^{\rightarrow}$ to a signal $z_1 : H_j \to \B$ and $y_2$ on $W_t^{\rightarrow}$ to a signal $z_2 : W_j \to \B$.
		Then, $\pi_{H_j}(z_1) \in P_j$ and $\pi_{W_j}(z_2) \in Q_j$, and the timed-until witness condition has the same truth value.
		Therefore, $y(t) \in B_j$.
		Since $t$ was arbitrary, every letter of $\pi_{T_j}(y)$ belongs to $B_j$.

		It remains to bound the length of this value expression.
		If $T_j$ is a singleton, then $|\pi_{T_j}(y)| = 1 = N_j$.
		Otherwise, for every $c \in M_J$, the induction hypothesis and the definition of $E_{\psi_1}$ show that $E_{\psi_1}(T_j \oplus c)$ bounds the number of changes of $y_1$ on $(T_j \oplus c) \cap [0,d)$.
		Similarly, for every $c \in C_J$, we have that $E_{\psi_2}(T_j \oplus c)$ bounds the number of changes of $y_2$ on $(T_j \oplus c) \cap [0,d)$.

		We say that a time is a change point of a signal if the signal is not constant on any neighborhood of that time within its domain.
		Let $D$ be the number of times $t \in T_j$ such that either $t+c$ is a change point of $y_1$ for some $c \in M_J$, or $t+c$ is a change point of $y_2$ for some $c \in C_J$.
		It follows that $D \leq \sum_{c \in M_J} E_{\psi_1}(T_j \oplus c) + \sum_{c \in C_J} E_{\psi_2}(T_j \oplus c)$.
		Between two consecutive such times, no point $t+c$ with $c \in M_J$ passes a change point of $y_1$, and no point $t+c$ with $c \in C_J$ passes a change point of $y_2$.
		The truth value of the timed-until formula is therefore constant there.
		The $D$ times divide $T_j$ into at most $D+1$ open intervals and may each contribute one additional value.
		Consequently, $|\pi_{T_j}(y)| \leq 2D+1 \leq N_j$.
		Thus, $\pi_{T_j}(y) \in \Alt_{N_j}(B_j)$.

		Since the placement regions cover $I$ in temporal order, $\pi_I(y) = \destutter \left( \pi_{T_1}(y) \cdot \ldots \cdot \pi_{T_k}(y) \right)$.
		The symbolic output therefore includes $\pi_I(y)$.

		\emph{Let $\psi = \psi_1 \since_J \psi_2$ with $J \neq [0,\infty)$.}
		This case follows by the same argument with time reversed.
		For $t \in T'_j$, the map from $H_t^{\leftarrow}$ to $H'_j$ preserves the witness window and the condition that, for some $s \in W_t^{\leftarrow}$, $y_2(s) = 1$ and $y_1(r) = 1$ for every $r \in (s,t]$.
		Thus, $y(t) \in B'_j$.
		For the length bound, the moving points for $y_1$ are $t-c$ with $c \in M_J$, and those for $y_2$ are $t-c$ with $c \in C_J$.
		For each $c$, these points range over $T'_j \ominus c$.
		Since $0 \in M_J$, the first group includes the evaluation point $t$ and therefore accounts for changes of $y_1$ there.
	\end{proof}

	\begin{proof}[\normalsize Proof of \cref{cl:algo}.]
		\normalsize
		Suppose first that $\varphi$ is a future formula, and let $I_0$ be the first segment of $G_S$.
		For every $w \in \tr(S,{\hb})$, \cref{cl:trsound} gives $w \in \tr^+(S,{\hb})$, and \cref{cl:evalsound} gives $\pi_{I_0}(y_{\varphi,w}) \in \llbracket (S,{\hb})\models \varphi \rrbracket$.
		Therefore, if all first letters in the symbolic output are $1$, then every trace in $\tr(S,{\hb})$ satisfies $\varphi$, and $[(S,{\hb})\models\varphi] = \top$.
		If all first letters are $0$, then every trace violates $\varphi$, and $[(S,{\hb})\models\varphi] = \bot$.
		For a past formula, let $I_f$ be the last segment of $G_S$ and use the same argument with last letter of $\pi_{I_f}(y_{\varphi,w})$.
	\end{proof}

	\paragraph*{Sets of Boolean Value Expressions as Bit Vectors.}
	Explicit asynchronous products are expensive to compute.
	Our implementation instead stores sets of Boolean value expressions compactly using bit vectors indexed by the first bit and the word length.
	Indeed, every Boolean value expression is destuttered and thus uniquely determined by these two pieces of information.
	We use this structure to avoid enumerating all pairs of value expressions.
	More precisely, when combining two sets, we handle the singleton expressions $0$ and $1$ separately.
	For the non-singleton expressions, the longest expression of each of the four endpoint types $0\ldots 0$, $0\ldots 1$, $1\ldots 0$, and $1\ldots 1$ in each set determines the longest possible output expression of each endpoint type.
	In most cases, every shorter output expression of the same endpoint type is also possible, so we add all of them at once.
	In the few cases where this is not true, we check the stored input bits to determine exactly which shorter output expressions are possible.

	\paragraph*{Numerical Abstractions for Real-Valued Signals.}
	Our monitoring algorithm, denoted \textsc{Adm}, supports piecewise-constant real-valued signals and numerical predicates.
	For a finite input, each component signal takes only finitely many values, which form the alphabet of its value expressions.
	The atomic rule aligns value expressions for each component and evaluates the numerical predicate pointwise.
	For example, for $x_1 + x_2 > 2$, aligning $2 \cdot 2 \cdot 3$ with $1 \cdot 0 \cdot 1$ gives the pointwise sum $3 \cdot 2 \cdot 4$ and then the Boolean value expression $101$.

	For some formula patterns, we use cheaper summaries instead.
	The \emph{Fine} abstraction, denoted \textsc{Adm-F}, ignores the order in which values occur within a segment.
	For each signal $x_i$, it keeps only the set $X_i$ of values that may appear in its value expressions.
	It then applies the numerical operation to every combination of values from these sets.
	For instance, the example above gives $X_1 = \{2,3\}$ and $X_2 = \{0,1\}$, and the possible sums are $\{2,3,4\}$.
	This includes every sum that can actually occur, since the values of the signals at any given time form one of the combinations considered.
	This approach can be less precise for formulas that depend on this order, such as $(x_1 > c_1) \until (x_2 > c_2)$, but it remains useful for formulas that do not, such as $\LTLg (x_1 + x_2 > c)$.

	The \emph{Coarse} abstraction, denoted \textsc{Adm-C}, summarizes \textsc{Adm-F} further by keeping only the extreme values.
	If the numerical expression is monotone in each argument, its output extrema are attained at the appropriate input extrema, so they preserve the possible outcomes of a threshold comparison.
	In the example above, \textsc{Adm-F} obtains the set $\{2,3,4\}$ of sums, whereas \textsc{Adm-C} keeps only its minimum $2$ and maximum $4$.
	
	We use \textsc{Adm-F} only for formula patterns that are insensitive to the order of values within a segment, and \textsc{Adm-C} only when the relevant numerical expressions are also monotone.
	In these cases, every conclusive verdict is sound.

	So far, we have assumed that the central monitor runs on a process distinct from the observed agents.
	If the monitor runs on one of the observed agents, we can instead use that agent's local clock as its reference.
	This removes the clock uncertainty of that agent relative to the monitor and can reduce the number of potential interleavings that the monitor needs to consider.
	We call this approach \emph{Relative}, denoted \textsc{Adm-Fr} when combined with \textsc{Adm-F} and \textsc{Adm-Cr} when combined with \textsc{Adm-C}.
	We use this variant only for untimed formulas; for timed formulas, it would use a semantics in which time is measured by the monitor agent's clock.
	We evaluate these in \cref{sec:experiments}.

	\paragraph*{Combining Exact and Approximate Monitoring.}
	We also use \textsc{Adm} as a filter before running the exact monitor \textsc{Edm}.
	The goal is to skip the exact computation whenever \textsc{Adm} is already conclusive, without changing the final verdict.

	Given a distributed signal $(S,{\hb})$ and a formula $\varphi$, we first compute $\llbracket (S,{\hb}) \models \varphi \rrbracket$.
	If the obtained verdict is conclusive (i.e., $\top$ or $\bot$), we return it.
	Otherwise, if it is ${?}$, we run \textsc{Edm} and return the exact verdict $[(S,{\hb}) \models \varphi]$.
	Since \textsc{Adm} is sound, the combined procedure returns the same three-valued verdict as \textsc{Edm}, while avoiding some calls to it.
	We evaluate this combination in \cref{sec:experiments}.

	\subsection{Online Algorithm}
	\label{sec:online}

	We now consider the online monitoring of a past STL formula $\varphi$ over a distributed signal $(S,{\hb})$.
	For simplicity, we assume that the input arrives in \emph{portions} of length $\Delta$.
	The $i$th portion contains the observations whose local timestamps lie in $[(i-1)\Delta,i\Delta)$.
	After this portion, the monitor has received all observations with local timestamps in $[0,i\Delta)$.
	We define $d_i \defeq i\Delta$ for all $i \geq 0$ and write $D_i \defeq (S^i,{\hb}^i)$ for this observed prefix.

	Throughout this subsection, we allow the untimed operator $\since = \since_{[0,\infty)}$, but assume that the intervals in timed operators are bounded from above.
	We explain at the end how the remaining unbounded intervals can be rewritten before monitoring.
	Moreover, we assume that $(S,{\hb})$ is a distributed signal observed incrementally in portions and $\varphi$ is a past STL formula.

	\paragraph*{Finalized Prefix.}
	Because of the clock-skew bound $\varepsilon$, later portions may still affect the most recent part of $D_i$.
	Indeed, if an edge received later has a local timestamp $t \in [d_i, d_i + \varepsilon)$, its uncertainty region may intersect with the preceding segment.
	We therefore set $q_i \defeq \max(0,d_i-\varepsilon)$ for all $i \geq 0$, and call $[0,q_i)$ the \emph{finalized prefix after portion $i$}.
	No edge received in later portions can occur in this prefix with respect to the reference clock; extending $D_i$ may extend an uncertainty region that was cut off at $d_i$, but only to the right of $d_i$.
	Therefore, later portions cannot change the canonical segmentation or the sets $\gamma(x,I)$ inside $[0,q_i)$.

	Since $q_i$ is not necessarily an endpoint in the canonical segmentation, we refine the canonical segmentation of $D_i$ by inserting the points $q_1, \ldots, q_i$.
	Let $G_i$ be the segments of this refined segmentation that are contained in $[0,q_i)$.
	If $q_i > 0$, let $K_i$ be the final segment of $G_i$, so $r_{K_i} = q_i$; if $q_i = 0$, let $G_i = \emptyset$.
	For every component signal $x$ and every refined segment $K \in G_i$, we compute $\gamma(x,K)$ from the observed prefix $D_i$ by the same rule as in \cref{sec:approach}.
	Then, we apply the evaluation rules from \cref{sec:offline} over the new segmentation $G_i$, taking $[0,q_i)$ as the temporal domain.
	The proofs of \cref{cl:trsound,cl:evalsound} apply to this refinement since every segment in $G_i$ is a subsegment of a canonical segment.
	For every subformula $\psi$ and every $K \in G_i$, let $\tau^i_{\psi,K}$ denote the resulting symbolic output.

	If $q_i > 0$, the monitor reads the last letters of $\tau^i_{\varphi,K_i}$.
	It returns $\top$ if all of them are $1$, $\bot$ if all are $0$, and ${?}$ otherwise.
	If $q_i = 0$, the monitor returns ${?}$ by default since the finalized prefix is empty.

	The \emph{naive} algorithm computes the symbolic outputs from scratch on the entire finalized prefix after every portion.
	The \emph{incremental} algorithm computes only the outputs on the newly finalized segments and keeps in memory only the earlier outputs that may still be needed.
	We illustrate the two approaches in \Cref{fig:online}.

	\begin{figure*}
		\centering
		\begin{subfigure}[c]{.1\textwidth}
			\centering
			\begin{tikzpicture}[scale=0.75,>=stealth]
				\node[anchor=east] at ($(-1,3.6)$) {};
				\vspace{1em}
				\node[anchor=east] at ($(-1,3)$) {\footnotesize{Portion 1}};
				\node[anchor=east] at ($(-1,1.5)$) {\footnotesize{Portion 2}};
				\node[anchor=east] at ($(-1,0)$) {{\footnotesize Portion 3}};
			\end{tikzpicture}
		\end{subfigure}
		\begin{subfigure}[c]{.42\textwidth}
			\centering
			\begin{tikzpicture}[
				scale=0.75,
				>=stealth,
				box/.style={draw=black},
				retained/.style={fill=gray!18},
				newportion/.style={fill=admAccentBlue!28}
			]
				\def\DL{2.5}
				\def\eps{1}
				\def\vsep{1.5}
				\def\arrowL{3*\DL+0.5}
				\def\barH{0.28}

				\coordinate (A) at (0,0);

				\scriptsize
				\node at ($(A)+(1.5*\DL,2.1*\vsep+0.9)$) {\textbf{(a)}};
				\footnotesize

				\draw[->] ($(A)+(0.5,2*\vsep)$) -- ++(\arrowL,0);
				\path ($(A)+(0.5,2*\vsep)$) coordinate (s1a);
				\fill[newportion] ($(s1a)+(0,-\barH/2)$) rectangle ++(\DL,\barH);
				\draw[box] ($(s1a)+(0,-\barH/2)$) rectangle ++(\DL,\barH);
				\draw[solid] ($(s1a)+(0,0.4)$) -- ++(0,-0.8);
				\draw[solid] ($(s1a)+(\DL,0.4)$) -- ++(0,-0.8);
				\draw[dashed] ($(s1a)+(\DL-\eps,0.4)$) -- ++(0,-0.8);
				\node[below] at ($(s1a)+(0,-0.4)$) {$0$};
				\node[below] at ($(s1a)+(\DL-\eps,-0.4)$) {$\Delta-\varepsilon$};
				\node[below] at ($(s1a)+(\DL,-0.4)$) {$\Delta$};

				\draw[->] ($(A)+(0.5,\vsep)$) -- ++(\arrowL,0);
				\path ($(A)+(0.5,\vsep)$) coordinate (s2a);
				\fill[retained] ($(s2a)+(0,-\barH/2)$) rectangle ++(2*\DL,\barH);
				\fill[newportion] ($(s2a)+(\DL,-\barH/2)$) rectangle ++(\DL,\barH);
				\draw[box] ($(s2a)+(0,-\barH/2)$) rectangle ++(2*\DL,\barH);
				\draw[dashed] ($(s2a)+(2*\DL-\eps,0.4)$) -- ++(0,-0.8);
				\draw[solid] ($(s2a)+(0,0.4)$) -- ++(0,-0.8);
				\draw[solid] ($(s2a)+(2*\DL,0.4)$) -- ++(0,-0.8);
				\node[below] at ($(s2a)+(0,-0.4)$) {$0$};
				\node[below] at ($(s2a)+(2*\DL-\eps,-0.4)$) {$2\Delta-\varepsilon$};
				\node[below] at ($(s2a)+(2*\DL,-0.4)$) {$2\Delta$};

				\draw[->] ($(A)+(0.5,0)$) -- ++(\arrowL,0);
				\path ($(A)+(0.5,0)$) coordinate (s3a);
				\fill[retained] ($(s3a)+(0,-\barH/2)$) rectangle ++(3*\DL,\barH);
				\fill[newportion] ($(s3a)+(2*\DL,-\barH/2)$) rectangle ++(\DL,\barH);
				\draw[box] ($(s3a)+(0,-\barH/2)$) rectangle ++(3*\DL,\barH);
				\draw[dashed] ($(s3a)+(3*\DL-\eps,0.4)$) -- ++(0,-0.8);
				\draw[solid] ($(s3a)+(0,0.4)$) -- ++(0,-0.8);
				\draw[solid] ($(s3a)+(3*\DL,0.4)$) -- ++(0,-0.8);
				\node[below] at ($(s3a)+(0,-0.4)$) {$0$};
				\node[below] at ($(s3a)+(3*\DL-\eps,-0.4)$) {$3\Delta-\varepsilon$};
				\node[below] at ($(s3a)+(3*\DL,-0.4)$) {$3\Delta$};
			\end{tikzpicture}
		\end{subfigure}
		\hspace{1em}
		\begin{subfigure}[c]{.42\textwidth}
			\centering
			\begin{tikzpicture}[
				scale=0.75,
				>=stealth,
				box/.style={draw=black},
				retained/.style={fill=gray!18},
				newportion/.style={fill=admAccentBlue!28}
			]
				\def\DL{2.5}
				\def\eps{1}
				\def\vsep{1.5}
				\def\arrowL{3*\DL+0.5}
				\def\barH{0.28}

				\coordinate (B) at (0,0);

				\scriptsize
				\node at ($(B)+(1.5*\DL,2.1*\vsep+0.9)$) {\textbf{(b)}};
				\footnotesize

				\draw[->] ($(B)+(0.5,2*\vsep)$) -- ++(\arrowL,0);
				\path ($(B)+(0.5,2*\vsep)$) coordinate (s1b);
				\fill[newportion] ($(s1b)+(0,-\barH/2)$) rectangle ++(\DL,\barH);
				\draw[box] ($(s1b)+(0,-\barH/2)$) rectangle ++(\DL,\barH);
				\draw[dashed] ($(s1b)+(\DL-\eps,0.4)$) -- ++(0,-0.8);
				\draw[solid] ($(s1b)+(0,0.4)$) -- ++(0,-0.8);
				\draw[solid] ($(s1b)+(\DL,0.4)$) -- ++(0,-0.8);
				\node[below] at ($(s1b)+(0,-0.4)$) {$0$};
				\node[below] at ($(s1b)+(\DL-\eps,-0.4)$) {$\Delta-\varepsilon$};
				\node[below] at ($(s1b)+(\DL,-0.4)$) {$\Delta$};

				\draw[->] ($(B)+(0.5,\vsep)$) -- ++(\arrowL,0);
				\path ($(B)+(0.5+\DL-2*\eps,\vsep)$) coordinate (s2b);
				\fill[retained] ($(s2b)+(0,-\barH/2)$) rectangle ++(\DL+2*\eps,\barH);
				\fill[newportion] ($(s2b)+(2*\eps,-\barH/2)$) rectangle ++(\DL,\barH);
				\draw[box] ($(s2b)+(0,-\barH/2)$) rectangle ++(\DL+2*\eps,\barH);
				\draw[dashed] ($(s2b)+(\DL+\eps,0.4)$) -- ++(0,-0.8);
				\draw[solid] ($(s2b)+(0,0.4)$) -- ++(0,-0.8);
				\draw[solid] ($(s2b)+(\DL+2*\eps,0.4)$) -- ++(0,-0.8);
				\node[below] at ($(s2b)+(0,-0.4)$) {$\Delta-2\varepsilon$};
				\node[below] at ($(s2b)+(\DL+\eps,-0.4)$) {$2\Delta-\varepsilon$};
				\node[below] at ($(s2b)+(\DL+2*\eps,-0.4)$) {$2\Delta$};

				\draw[->] ($(B)+(0.5,0)$) -- ++(\arrowL,0);
				\path ($(B)+(0.5+2*\DL-2*\eps,0)$) coordinate (s3b);
				\fill[retained] ($(s3b)+(0,-\barH/2)$) rectangle ++(\DL+2*\eps,\barH);
				\fill[newportion] ($(s3b)+(2*\eps,-\barH/2)$) rectangle ++(\DL,\barH);
				\draw[box] ($(s3b)+(0,-\barH/2)$) rectangle ++(\DL+2*\eps,\barH);
				\draw[dashed] ($(s3b)+(\DL+\eps,0.4)$) -- ++(0,-0.8);
				\draw[solid] ($(s3b)+(0,0.4)$) -- ++(0,-0.8);
				\draw[solid] ($(s3b)+(\DL+2*\eps,0.4)$) -- ++(0,-0.8);
				\node[below] at ($(s3b)+(0,-0.4)$) {$2\Delta-2\varepsilon$};
				\node[below] at ($(s3b)+(\DL+\eps,-0.4)$) {$3\Delta-\varepsilon$};
				\node[below] at ($(s3b)+(\DL+2*\eps,-0.4)$) {$3\Delta$};
			\end{tikzpicture}
		\end{subfigure}
		\vspace{1ex}
		\caption{
			Online monitoring with portion length $\Delta$ and clock-skew bound $\varepsilon$.
			The horizontal arrows denote time, the light gray regions show the observations retained from the previous portions, the muted blue regions show the new portions, and the dashed lines show the finalized time $q_i$.
			\textbf{(a)} \textsc{Adm-Naive} keeps the entire observed prefix and computes the symbolic outputs for the entire prefix from scratch after each new portion.
			\textbf{(b)} \textsc{Adm-Incr} keeps only the new portion together with the suffix retained from the previous portion, which starts at $\max(0,d_{i-1}-2\varepsilon)$.
			Note that the drawing assumes $\Delta \geq 2\varepsilon$.
			\label{fig:online}
		}
	\end{figure*}

	\paragraph*{Naive Evaluation.}
	The reference algorithm, denoted \textsc{Adm-Naive}, keeps the entire observed prefix and proceeds as follows after receiving the $i$th portion:
	\begin{enumerate}[label=\arabic*.]
		\item Append the new portion and obtain $D_i$.
		\item Compute $q_i$ and the refined segmentation $G_i$ of the finalized prefix.
		\item If $q_i = 0$, return ${?}$.
		Otherwise, process the segments of $G_i$ from left to right and, on each segment $K$, compute $\tau^i_{\psi,K}$ for all subformulas $\psi$ of $\varphi$ in bottom-up order, using the offline symbolic rules from \cref{sec:offline}.
		\item Let $\tau_i^{\mathrm{nai}} \defeq \tau^i_{\varphi,K_i}$ and return the verdict obtained from $\last(\tau_i^{\mathrm{nai}})$.
	\end{enumerate}

	The naive algorithm is simple and its soundness follows immediately from that of offline evaluation, but it is inefficient since its running time and memory use grow with the observed prefix.

	\begin{theorem} \label{cl:online-sound-naive}
		For every $i \geq 1$ with $q_i > 0$
		\[
			\{\pi_{K_i}(y_{\varphi,w}) \st w \in \tr(S,{\hb})\}
			\subseteq
			\tau_i^{\mathrm{nai}},
		\]
		and therefore
		\begin{align*}
			\last(\tau_i^{\mathrm{nai}})=\{1\}
			&\Rightarrow \forall w \in \tr(S,{\hb}) : y_{\varphi,w}(q_i^-) = 1,\\
			\last(\tau_i^{\mathrm{nai}})=\{0\}
			&\Rightarrow \forall w \in \tr(S,{\hb}) : y_{\varphi,w}(q_i^-) = 0.
		\end{align*}
	\end{theorem}
	\begin{proof}[\normalsize Proof.]
		\normalsize
		Let $w \in \tr(S,{\hb})$ and consider its restriction to $[0,q_i)$.
		Every edge received after portion $i$ has local timestamp at least $d_i$, so its uncertainty region starts at or after $q_i$ and it cannot occur in this restriction.
		Moreover, extending $D_i$ changes no uncertainty region inside $[0,q_i)$.
		Since no later portion can add an event before $q_i$, the signal $D_i$ already contains all information relevant to every segment of $G_i$.
		Therefore, by the same argument as in \cref{cl:trsound}, the behavior of $w$ on each segment is included in the corresponding $\gamma$-sets used by the naive evaluator.
		Since $\varphi$ is a past formula, its satisfaction signal before $q_i$ depends only on this restriction.
		Applying the segmentwise argument of \cref{cl:evalsound} to the refined segmentation $G_i$ therefore gives $\pi_{K_i}(y_{\varphi,w}) \in \tau_i^{\mathrm{nai}}$.
		Since $w$ was arbitrary, the stated inclusion follows, and the claims about the last letters are immediate.
	\end{proof}

	\paragraph*{Incremental Evaluation.}
	The incremental algorithm \textsc{Adm-Incr} keeps only (i) the input data needed to construct new segments and their value expressions and (ii) the previously computed symbolic outputs that a past operator may need.
	
	First, we determine how much input data \textsc{Adm-Incr} must keep in memory.
	Notice that an edge with $t + \varepsilon \leq q_i$ cannot affect any segment after portion $i$ (because the left endpoint of every such segment is at least $q_i$), so the monitor does not need to retain such edges.
	We therefore define the input cutoff after portion $i$ as $b_i \defeq \max(0, q_i - \varepsilon) = \max(0, d_i - 2\varepsilon)$ for every $i \geq 0$.
	After portion $i$, the monitor keeps, for each component signal $x$, the value $x(b_i)$ and every edge with local timestamp greater than $b_i$, and discards all other edges.

	Now, we determine how much of the previously computed symbolic output \textsc{Adm-Incr} must retain.
	For untimed formulas, only the rule for $\since$ refers to an earlier segment, and it uses only the last letters of the immediately preceding segment.
	A timed formula such as $\psi_1 \since_J \psi_2$, on the other hand, may read the outputs of its subformulas on several earlier segments.
	We therefore define the history bound of $\varphi$ as
	\[
		H_\varphi \defeq \max \left( \{0\} \cup \left\{ \sup J \;\middle|\; \substack{ \psi_1 \since_J \psi_2 \text{ occurs in } \varphi,\\ \sup J < \infty } \right\} \right).
	\]
	After portion $i$ has been processed, every time evaluated in a later round satisfies $t \geq q_i$.
	If a bounded timed formula $\psi_1 \since_J \psi_2$ is evaluated in a later round at time $t$, then
	\[
		t - \sup J \geq q_i - \sup J \geq q_i - H_\varphi.
	\]
	Therefore, every previously finalized segment whose output may still be needed has right endpoint at least $q_i - H_\varphi$.

	Whenever the incremental algorithm computes the symbolic output of a subformula $\psi$ on a segment $K$, we denote it by $\tau^{\mathrm{inc}}_{\psi,K}$.
	After portion~$i$, the monitor retains every segment $K \in G_i$ with $r_K \geq q_i - H_\varphi$, together with $\tau^{\mathrm{inc}}_{\psi,K}$ for every subformula $\psi$ of $\varphi$.
	Moreover, since $r_{K_i} = q_i$, it always retains $K_i$ when $q_i > 0$.

	Starting with $q_0 = 0$ and no retained symbolic outputs, \textsc{Adm-Incr} proceeds as follows after receiving portion $i$.
	\begin{enumerate}[label=\arabic*.]
		\item Append the new portion to the retained input data and compute $q_i$.
		\item Use this input data to construct the segments of $G_i$ contained in $[q_{i-1},q_i)$, inserting $q_{i-1}$ and $q_i$ as endpoints.
		\item Process these segments from left to right and evaluate the subformulas of $\varphi$ bottom-up, using the offline symbolic rules.
		\item If $q_i = 0$, set $v_i \defeq {?}$.
		Otherwise, let $\tau_i^{\mathrm{inc}} \defeq \tau^{\mathrm{inc}}_{\varphi,K_i}$ be the output computed in this round, and let $v_i$ be the verdict obtained from $\last(\tau_i^{\mathrm{inc}})$.
		\item Discard every retained segment $K$ with $r_K < q_i - H_\varphi$, together with all symbolic outputs stored on it.
		For each component signal $x$, keep only its value at $b_i$ and the edges with local timestamp greater than $b_i$.
		\item Return $v_i$.
	\end{enumerate}
	
	The retained input data is used only to construct the new segments and their component value expressions, and the retained symbolic outputs are used only when a past operator reaches to the left of $q_{i-1}$.
	So, no formula is evaluated again on a segment finalized in an earlier round.
	After pruning, the retained input spans at most $2\varepsilon$ units of local time, and the retained symbolic outputs lie in the last $H_\varphi$ units of the finalized prefix, apart from at most one segment ending at or crossing the cutoff.

	\begin{theorem}\label{cl:algoOnline}
		For every $i \geq 1$ with $q_i > 0$, we have $\tau_i^{\mathrm{nai}}=\tau_i^{\mathrm{inc}}$, i.e., \textsc{Adm-Naive} and \textsc{Adm-Incr} return the same verdict after every portion.
	\end{theorem}
	\begin{proof}[\normalsize Proof.]
		\normalsize
		We prove the statement by induction over portions.
		After processing and pruning portion $i$, we maintain the following two invariants:
		\begin{enumerate}
			\item For every subformula $\psi$ of $\varphi$ and every segment $K \in G_i$ with $r_K \geq q_i-H_\varphi$, the stored output agrees with the output computed from the entire observed prefix $D_i$, i.e., $\tau^{\mathrm{inc}}_{\psi,K} = \tau^i_{\psi,K}$.
			\item On $[q_i,d_i)$, the stored input induces the same segment endpoints as the entire observed prefix $D_i$ and, for every component signal $x$ and every such segment $K$, the same set $\gamma(x,K)$.
		\end{enumerate}
		Both invariants hold before any input has been received.

		Assume that they hold after portion $i-1$, and consider portion $i$.
		Every edge discarded after portion $i-1$ has a local timestamp $t \leq b_{i-1}$.
		If $b_{i-1} > 0$, then $b_{i-1} = q_{i-1} - \varepsilon$, and therefore $t + \varepsilon \leq q_{i-1}$.
		If $b_{i-1} = 0$, no edge was discarded.
		Thus, every discarded uncertainty region ends at or before $q_{i-1}$.

		Every edge in portion $i$ has a local timestamp $t \geq d_{i-1}$, so the left endpoint of its uncertainty region is at least $\max(0,d_{i-1}-\varepsilon) = q_{i-1}$.
		Moreover, every earlier edge whose uncertainty region can extend beyond $d_{i-1}$ is still retained.
		This is immediate when $b_{i-1} = 0$; otherwise, $t + \varepsilon > d_{i-1}$ implies $t > b_{i-1}$.
		Together with the stored value of each component signal at $b_{i-1}$, the retained input therefore produces the same new segments and the same sets $\gamma(x,K)$ as $D_i$ between $q_{i-1}$ and $q_i$.
		The segments before $q_{i-1}$ are already finalized, so adding portion $i$ cannot change their component value expressions or the symbolic outputs of past formulas on them.

		We now compare the symbolic outputs on the new segments.
		We proceed from left to right and, on each segment, by induction on the formula.
		The cases of atomic propositions and Boolean operators are immediate from the equality of the corresponding $\gamma$ sets and the symbolic outputs of their immediate subformulas.

		For untimed since, the evaluation rule also uses the last-letter sets of the since formula and its left subformula on the preceding segment.
		For the first new segment, these sets are stored on $K_{i-1}$ and are correct by the induction hypothesis when $q_{i-1} > 0$.
		When $q_{i-1} = 0$, both algorithms use the same convention.
		For every later new segment, the required sets come from the preceding segment computed in the current round.

		Now, consider a bounded formula $\psi_1 \since_J \psi_2$ evaluated at a time $t$ on a new segment.
		Since $t \geq q_{i-1}$ and $\sup J \leq H_\varphi$, we have $t - \sup J \geq q_{i-1} - H_\varphi$.
		So, every previously finalized segment that the horizon has a nonempty intersection with was retained after portion $i-1$; all other segments it has a nonempty intersection with are computed in the current round.
		The outputs for $\psi_1$ and $\psi_2$ agree on the retained segments by the induction over portions and on the new segments by the induction over segments and the induction on the formula.
		The timed-since rule therefore gives the same output in both algorithms.
		We have now covered every formula case on every new segment, so the two algorithms provide the same outputs in portion $i$.

		We complete the proof by showing that pruning preserves the two invariants.
		For the first invariant, consider a segment $K$ finalized before portion $i$ and retained after pruning.
		Then, $r_K \geq q_i - H_\varphi$.
		Since $q_i \geq q_{i-1}$, we also have $r_K \geq q_{i-1} - H_\varphi$, so $K$ was retained after portion $i-1$ and its stored output was correct by the induction hypothesis.
		Because $K$ was already finalized, this output does not change when portion $i$ is added.
		The outputs retained on segments finalized in portion $i$ are correct by the argument above.
		Thus, the first invariant is preserved.

		For the second invariant, every edge discarded after portion $i$ has an uncertainty region ending at or before $q_i$, while the stored value $x(b_i)$ and the remaining edges reconstruct each component signal from $b_i$ onward.
		Hence, the stored input induces the same segment endpoints and the same sets $\gamma(x,K)$ as $D_i$ on $[q_i,d_i)$.
		The second invariant is therefore also preserved.

		If $q_i = 0$, both algorithms return ${?}$.
		Otherwise, $K_i$ is among the segments finalized in portion $i$, so the equality proved above gives $\tau_i^{\mathrm{inc}} = \tau_i^{\mathrm{nai}}$.
	\end{proof}

	\begin{corollary}\label{cl:online-sound}
		For every $i \geq 1$ with $q_i > 0$
		\[
			\{\pi_{K_i}(y_{\varphi,w}) \st w \in \tr(S,{\hb})\}
			\subseteq
			\tau_i^{\mathrm{inc}},
		\]
		and therefore
		\begin{align*}
			\last(\tau_i^{\mathrm{inc}})=\{1\}
			&\Rightarrow \forall w \in \tr(S,{\hb}) : y_{\varphi,w}(q_i^-) = 1,\\
			\last(\tau_i^{\mathrm{inc}})=\{0\}
			&\Rightarrow \forall w \in \tr(S,{\hb}) : y_{\varphi,w}(q_i^-) = 0.
		\end{align*}
	\end{corollary}
	\begin{proof}[\normalsize Proof.]
		\normalsize
		Follows from \cref{cl:online-sound-naive,cl:algoOnline}.
	\end{proof}

	\paragraph*{Remark on Unbounded Timing Intervals.}
	We have assumed above that every timing interval is either $[0,\infty)$ or bounded from above.
	The remaining unbounded intervals can be removed before monitoring by rewriting the formula as a combination of untimed operators and bounded timed operators.

	\subparagraph*{Closed Lower Endpoint}
	Consider the formula $\psi_1 \since_{[a,\infty)} \psi_2$ where $a > 0$.
	A direct rewriting gives
	\[
		\psi_1 \since_{[a,\infty)} \psi_2
		\equiv
		\psi_1 \since_{\{a\}} (\psi_1 \since \psi_2).
	\]
	At time $t$, the inner untimed formula chooses a witness in $[0,t-a]$, while the outer formula checks that $\psi_1$ holds throughout $(t-a,t]$.
	The timed evaluation rule handles the (singleton) interval $\{a\}$ as described above for arbitrary intervals.

	Alternatively, for any finite $b > a$, we have
	\[
		\psi_1 \since_{[a,\infty)} \psi_2
		\equiv
		\psi_1 \since_{[a,b]} (\psi_1 \since \psi_2).
	\]
	At time $t$, the outer formula may choose any point between $t-b$ and $t-a$, and the inner untimed formula summarizes everything before that point.

	The first rewriting is usually preferable for the symbolic monitor since it needs only $a$ time units of earlier symbolic outputs and uses the fixed point $t-a$.
	The second needs $b$ time units and introduces the additional offset $b$, which can create more placement regions and increase the length bounds, which adds further overapproximation.

	\subparagraph*{Open Lower Endpoint}
	Now, consider the formula $\psi_1 \since_{(a,\infty)} \psi_2$ where $a \geq 0$.
	We cannot split the formula exactly at $t-a$ as above, since a witness at distance $a$ is not allowed by the formula.

	Consider a finite $b > a$.
	Then,
	\[
		\psi_1 \since_{(a,\infty)} \psi_2
		\equiv
		\psi_1 \since_{(a,b]} (\psi_1 \since \psi_2).
	\]
	The outer formula first chooses a point $\hat{t}$ in $[t-b,t-a)$.
	The inner formula then chooses a witness in $[0,\hat{t}]$, and the outer formula checks $\psi_1$ throughout $(\hat{t}, t]$.
	Together, they cover exactly the witnesses whose distance from $t$ is greater than $a$.

	Equivalently, we may separate the nearer and more distant witnesses:
	\[
		\psi_1 \since_{(a,\infty)} \psi_2
		\equiv
		(\psi_1 \since_{(a,b)} \psi_2)
		\lor
		(\psi_1 \since_{\{b\}} (\psi_1 \since \psi_2)).
	\]
	The first disjunct covers distances in $(a,b)$, and the second covers distances in $[b,\infty)$.

	Both rewritings contribute $b$ to the history bound $H_\varphi$, but the first uses only one bounded timed operator and is therefore simpler.

	\subparagraph*{Rewriting Precision.}
	All four rewritings preserve the standard STL semantics.
	Their symbolic outputs need not coincide because the evaluator is sensitive to syntax and combines subformula outputs independently.
	Nevertheless, using our algorithm with the rewritten formulas is sound for the original unbounded formula by the semantic equivalence and \cref{cl:online-sound}.

	\section{Experimental Evaluation} 
	\label{sec:experiments}
	
	We seek answers to the following research questions (RQs):
	\begin{resq}[What is the tradeoff between the efficiency and the conclusiveness of approximate distributed monitors?]
		Approximate distributed monitoring trades speed for conclusiveness: relative to the exact monitor, it may return $?$ on instances where the exact verdict is already $\top$ or $\bot$.
		We want to understand the tradeoff between the potential speedups that an approximate distributed monitor can achieve when compared to its exact counterpart and the corresponding loss of conclusiveness due to the approximations.
		We would also like to identify the classes of signals and properties for which this tradeoff is effective. 
	\end{resq}
	\begin{resq}[Can the combination of approximate and exact distributed monitors increase efficiency while preserving precision?]
		We are interested in evaluating whether a combined use of approximate and exact distributed monitors can improve monitoring efficiency while preserving the exact three-valued verdicts.
	\end{resq}
	\begin{resq}[How scalable are online incremental monitors?]
		The offline monitoring algorithm can be naively used for online monitoring, but repeatedly processing the entire observed prefix incurs increasing time and memory costs.
		The incremental approach instead evaluates newly finalized segments and retains only the bounded history needed by future updates.
		We evaluate its runtime scaling with trace duration, observation frequency, and clock skew.
	\end{resq}

	\begin{table*}[t]
		\centering
			\begin{tabular}{|l|l|}
				\hline
				Subject & STL formula(s) \\
				\hline
				RG (offline) & $\varphi_1 \defeq \LTLg (p \wedge q)$ \,\,\,\,\,\,\,\,\,\,\,\,\,\,\, $\varphi_2 \defeq \LTLg (p \lor q)$ \,\,\,\,\,\,\,\,\,\,\,\,\,\,\,\,\,\,\,\,\,\,\,\,\, $\varphi_3 \defeq p \until q$ \\
				& $\varphi_4 \defeq \LTLg (p \Rightarrow \LTLf q)$  \,\,\,\,\,\,  $\varphi_5 \defeq \LTLg (p \Rightarrow \LTLf_{[0,1)} q)$ \,\,\,\,\,\, $\varphi_6 \defeq \LTLg (p \Rightarrow \LTLf_{[0,2)} q)$ \\
				RG (online) & $\psi_1 \defeq \LTLsquareminus (p \wedge q)$ \,\,\,\,\,\,\,\,\,\,\,\,\,\,\, $\psi_2 \defeq \LTLsquareminus (p \lor q)$ \,\,\,\,\,\,\,\,\,\,\,\,\,\,\,\,\,\,\,\,\,\,\,\,\, $\psi_3 \defeq p \since q$ \\
				& $\psi_4 \defeq \LTLsquareminus (p \Rightarrow \LTLdiamondminus q)$  \,\,\,\,\,\,  $\psi_5 \defeq \LTLsquareminus (p \Rightarrow \LTLdiamondminus_{[0,1)} q)$ \,\,\,\,\,\, $\psi_6 \defeq \LTLsquareminus (p \Rightarrow \LTLdiamondminus_{[0,2)} q)$ \\
				WT & $\varphi_{\text{WT}} \defeq \LTLg \left(\sum_{i=1}^{n} x_i  > 10\right)$  \\
				SD & $\varphi_{\text{SD}} \defeq \bigwedge_{1 \leq i < j \leq n} \LTLg \left( \sqrt{(x_i-x_j)^2 + (y_i-y_j)^2 + (z_i-z_j)^2} > 0 \right)$   \\
				\hline
			\end{tabular}
		\caption{STL specifications used in the experiments.}
		\label{tab:spec} 
	\end{table*}

	\subsection{Experimental Setup}
	
	\paragraph*{Distributed Monitors.}
	In our study, we compare our approximate distributed monitoring (\textsc{Adm}) approach and its variants to an exact distributed monitoring approach (\textsc{Edm}).\footnote[1]{The code is available at \url{https://github.com/egesarac/ApxDistMon}.}
	For \textsc{Edm}, we use a variant of the distributed monitoring procedure from~\cite{MomtazAB23} that evaluates STL specifications over distributed traces via SMT solving.
	The SMT-based procedure originally assumes polynomial continuous signals.
	We adapt it to piecewise-constant signals so that both baselines operate under the same signal model as \textsc{Adm}.
	This adaptation makes the exact baseline slower than the implementation in~\cite{MomtazAB23}.
	Moreover, the exact monitor of~\cite{MomtazAB23} can operate incrementally by splitting the input trace into segments and potentially terminating early.
	Since our comparison here targets the offline algorithms, we use \textsc{Edm} in its non-incremental form.
	
	\paragraph*{Experimental Subjects.}
	To answer our research questions, we use (1) a \emph{random generator (RG)} of distributed traces, (2) a \emph{water tank (WT)} case study, and (3) a \emph{swarm of drones (SD)} case study.

	The \emph{Random Generator (RG)} uses a uniform distribution to generate distributed traces in which we control the trace duration $d$ and the skew bound $\varepsilon$ that determines the width of the uncertainty regions around event times.

	The \emph{Water Tank (WT)} subject is a Simulink model of a hybrid high-pressure water distribution system.
	In each benchmark instance, the distributed trace contains one outflow-pressure signal per tank, so $n$ denotes the number of water tanks and signals in that instance.
	The experiments vary $n$ to study scalability.
	Inlet pipes connect each tank to an external source, and outlet pipes distribute high-pressure water regulated by valves.
	Each valve is operated by a controller that samples the outflow pressure at 20\,Hz using its local clock.
	Our model is a simplified emulation of the Refueling Water Storage Tanks (RWST) module of an Emergency Core Cooling System (ECCS) of a Pressurized Water Reactor Plant~\cite{USNRCPWR}.

	The \emph{Swarm of Drones (SD)} subject is generated using the path-planning software Fly-by-Logic~\cite{PantAM17CCTA}.
	Here, a swarm of drones performs various reach-avoid missions while satisfying objectives such as reaching a goal within a deadline and avoiding obstacles and collisions.
	The path planner finds the most robust trajectory using a temporal-logic robustness optimizer.
	These trajectories are sampled at 20\,Hz.

	In these benchmarks, larger $\varepsilon$ means wider uncertainty regions; the configurations where $\varepsilon$ is comparable to or larger than the sampling interval are particularly challenging because samples from adjacent rounds can overlap in time.
	
	\paragraph*{Specifications.}
	Table~\ref{tab:spec} shows the STL specifications that we use to evaluate our experimental subjects.
	For offline monitoring, specifications $\varphi_{1}$ to $\varphi_{6}$ are monitored against the distributed traces created by the random generator and represent different classes and fragments of Boolean-valued temporal formulas.
	The specification $\varphi_1$ is a formula in which both the outer temporal operator ($\LTLg$) and the inner Boolean operator ($\wedge$) are conjunctive.
	The specification $\varphi_2$ combines conjunctive $(\LTLg)$ and disjunctive ($\lor$) operators, and $\varphi_3$ is the usual until operator.
	The remaining three formulas are the common response formulas, which also combine conjunctive $(\LTLg)$ and disjunctive $(\LTLf, \Rightarrow)$ operators: $\varphi_4$ is the standard response formula, $\varphi_5$ and $\varphi_6$ add a bounded real-time requirement to the previous specification, where the bound of $\varphi_5$ is tighter than that of $\varphi_6$.
	For online monitoring, we use the past-time variants $\psi_1$ to $\psi_6$ of the formulas described above.
	The specification $\varphi_{\text{WT}}$ associated to the water tank case study is an STL formula in which a sum of signals originating from different agents is compared to a constant.
	Finally, the specification $\varphi_{\text{SD}}$ defines a mutual separation property over a swarm of drones, requiring more sophisticated arithmetic operations on signals originating from different agents.

	\paragraph*{Computing Platform.}
	We ran the experiments on a laptop with Ubuntu 24.04, an Intel Core Ultra 7 255U CPU, and 32\,GB of RAM.
	\textsc{Adm} is implemented in C++ and compiled with \texttt{g++} 13.3.0 using \texttt{-O3}.
	\textsc{Edm}~\cite{MomtazAB23} is implemented in Python 3.12.3 and uses Z3 4.16.0~\cite{MouraB08}.

	Each RG parameter configuration uses 100 generated traces, while each case study (WT and SD) uses one fixed $1\,\mathrm{s}$ trace set sampled at $20\,\mathrm{Hz}$.
	For offline monitoring, we run \textsc{Adm} 100 times and \textsc{Edm} 3 times per trace.
	For online monitoring, we run both \textsc{Adm-Naive} and \textsc{Adm-Incr} 10 times per trace.
	We run each case study (WT and SD) configuration 10 times.
	All reported running times are averages over these repetitions and inputs.
	For \textsc{Edm}, we report the larger of the mean times for the positive and negative SMT checks; this is the time they would take if run in parallel.
	The timeout for each SMT check is 600 seconds for RG, 120 seconds for WT, and 360 seconds for SD.

	\subsection{Discussion}

\begin{figure*}[p]
		\begin{center}
			\includegraphics[width=\linewidth]{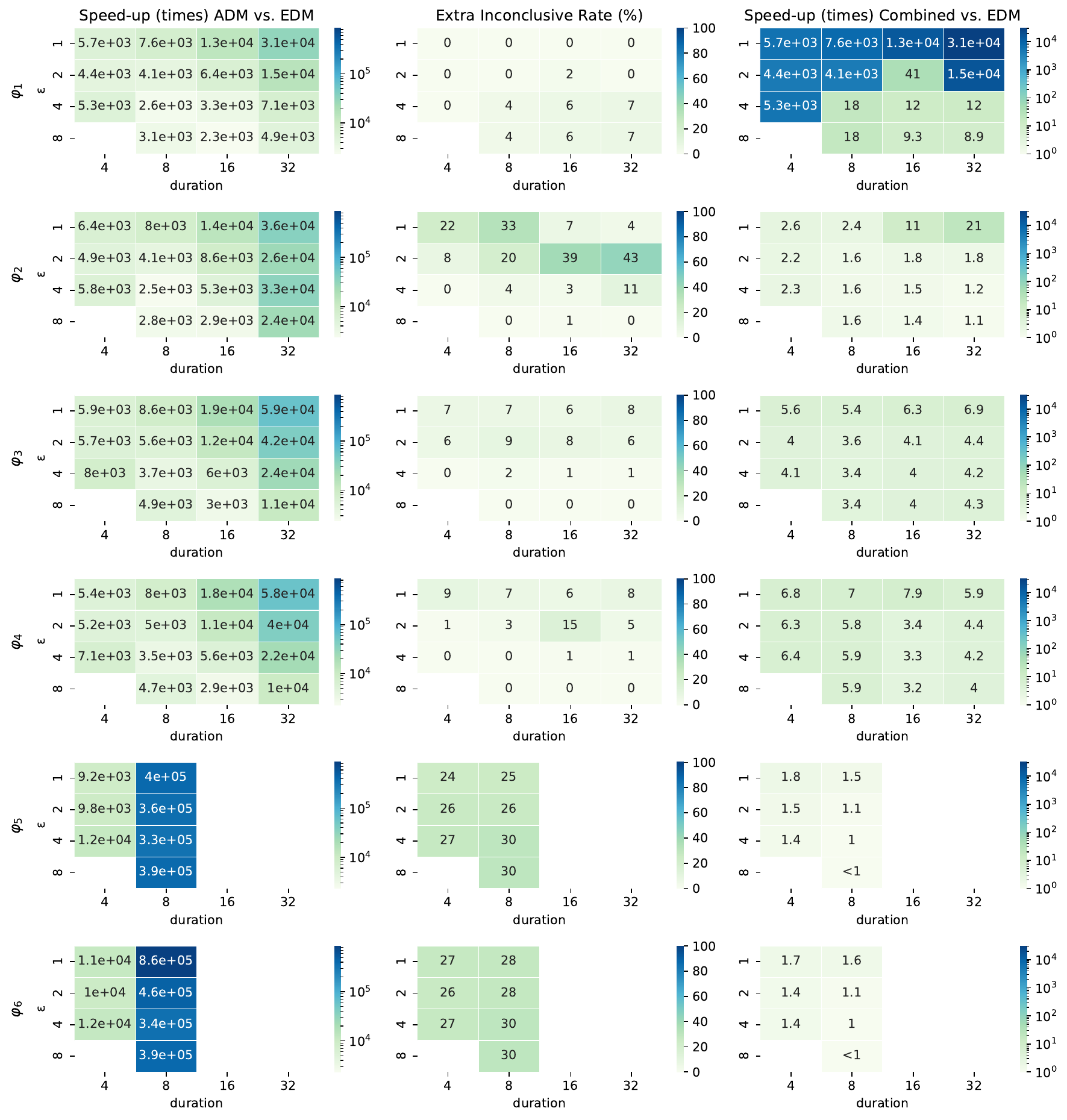}
			\caption{Results on offline monitoring $\varphi_{1}$ to $\varphi_{6}$ with \textsc{Adm} on distributed traces created by the RG.}
			\label{fig:rgresults}
		\end{center}
	\end{figure*}

	\paragraph*{Random Generator, Offline Monitoring.}
	Figure~\ref{fig:rgresults} summarizes the evaluation of specifications $\varphi_1$ to $\varphi_6$ on RG traces.
	The first column shows the speedup of \textsc{Adm} over \textsc{Edm} for each trace duration $d$ and clock skew $\varepsilon$.
	The second shows the percentage of instances where \textsc{Adm} returns $?$, but \textsc{Edm} returns $\top$ or $\bot$, which we call the \emph{extra inconclusive rate}.
	The third shows the estimated speedup over \textsc{Edm} of the combined monitor that first runs \textsc{Adm} and runs \textsc{Edm} only if \textsc{Adm} is inconclusive.

	The first column shows that \textsc{Adm} provides excellent speedup.
	Over the configurations without an \textsc{Edm} timeout, \textsc{Adm} is between $2{,}320\times$ and $397{,}000\times$ faster.
	When the timed out instances are included as $600\,\mathrm{s}$ (six instances for $\varphi_6$, five with $d = 8$ and $\varepsilon = 1$ and one with $d = 8$ and $\varepsilon = 2$), the speedup values go up to at least $856{,}000\times$.
	The mean running time of \textsc{Adm} is below $1.4\,\mathrm{ms}$ in every configuration.

	Considering the untimed formulas in the second column, \textsc{Adm} appears more precise on $\varphi_1$, $\varphi_3$, and $\varphi_4$ than on $\varphi_2$, with the extra inconclusive rate being between $0\%$ and $15\%$ for the former group and between $0\%$ and $43\%$ for the latter.
	However, most of this difference comes from simple cases that occur often in randomly generated traces.
	More specifically,
	$\varphi_1$ is false when $p(0) = 0$, $q(0) = 0$, $p(d^-) = 0$, or $q(d^-) = 0$, which happens in $1{,}423$ instances out of $1{,}500$;
	$\varphi_3$ is true when $q(0) = 1$ and it is false when $p(0) = q(0) = 0$, which happens in $1{,}129$ instances out of $1{,}500$; and
	$\varphi_4$ is true when $q(d^-) = 1$ and it is false when $p(d^-) = 1$ and $q(d^-) = 0$, which happens in $1{,}167$ instances out of $1{,}500$.
	On the other hand, $\varphi_2$ is false when $p(0) = q(0) = 0$ or $p(d^-) = q(d^-) = 0$, which happens only in $673$ instances out of $1{,}500$.
	Considering only the remaining, more difficult instances, the extra inconclusive rate of $\textsc{Adm}$ is $46.8\%$ for $\varphi_1$, $52.4\%$ for $\varphi_2$, $53\%$ for $\varphi_3$, and $53.8\%$ for $\varphi_4$.

	The timed formulas $\varphi_5$ and $\varphi_6$ introduce an additional loss in conclusiveness due to the profile construction.
	Combining potential placements of $q$ that cannot occur in one retiming may lead to a spurious response, causing \textsc{Adm} to return an inconclusive verdict on real violations.
	For $\varphi_5$, the extra inconclusive rate is $7\%$ for real satisfactions, while it is $68.9\%$ for violations.
	Similarly, for $\varphi_6$, the corresponding rates are $12.3\%$ and $75.2\%$, respectively.
	In particular, there are $207$ instances, out of $700$, with $p(d^-) = 1$ and $q(d^-) = 0$, which force a violation in both formulas, but \textsc{Adm} retains only $39$ of these for $\varphi_5$ and $33$ for $\varphi_6$.
	The longer timing interval of $\varphi_6$ increases the number of possible placements of $q$, which also explains its slightly higher inconclusive rates.

	The third column shows how the inconclusive outcomes affect the combined monitor.
	Across all $7{,}400$ instances, \textsc{Adm} returns $?$ on $32.9\%$ of the traces, but these traces account for $78.6\%$ of the total \textsc{Edm} time, so they contain most of the expensive \textsc{Edm} instances.
	This explains why a low extra inconclusive rate does not always give a large speedup for the combined monitor.
	For example, for $\varphi_2$ with $d = 32$ and $\varepsilon = 8$, the extra inconclusive rate is $0\%$, but both monitors return $?$ on $53\%$ of the traces, which account for $87.6\%$ of the \textsc{Edm} time, so the combined speedup is only $1.14\times$, despite a raw speedup of $23{,}800\times$.

	Overall, the combined monitor is estimated to be $1.27\times$ faster than \textsc{Edm}, reducing the total running time by $21.4\%$, and is faster in $72$ of the $74$ configurations.
	In the remaining two, corresponding to $\varphi_5$ and $\varphi_6$ with $d = 8$ and $\varepsilon = 8$, \textsc{Adm} is inconclusive on every trace, so the combined monitor is negligibly slower.
	Nonetheless, on three of the six instances where \textsc{Edm} timed out, \textsc{Adm} returns a conclusive verdict, so the combined monitor would avoid these timeouts.

\begin{figure*}[p]
		\begin{center}
			\includegraphics[width=\linewidth]{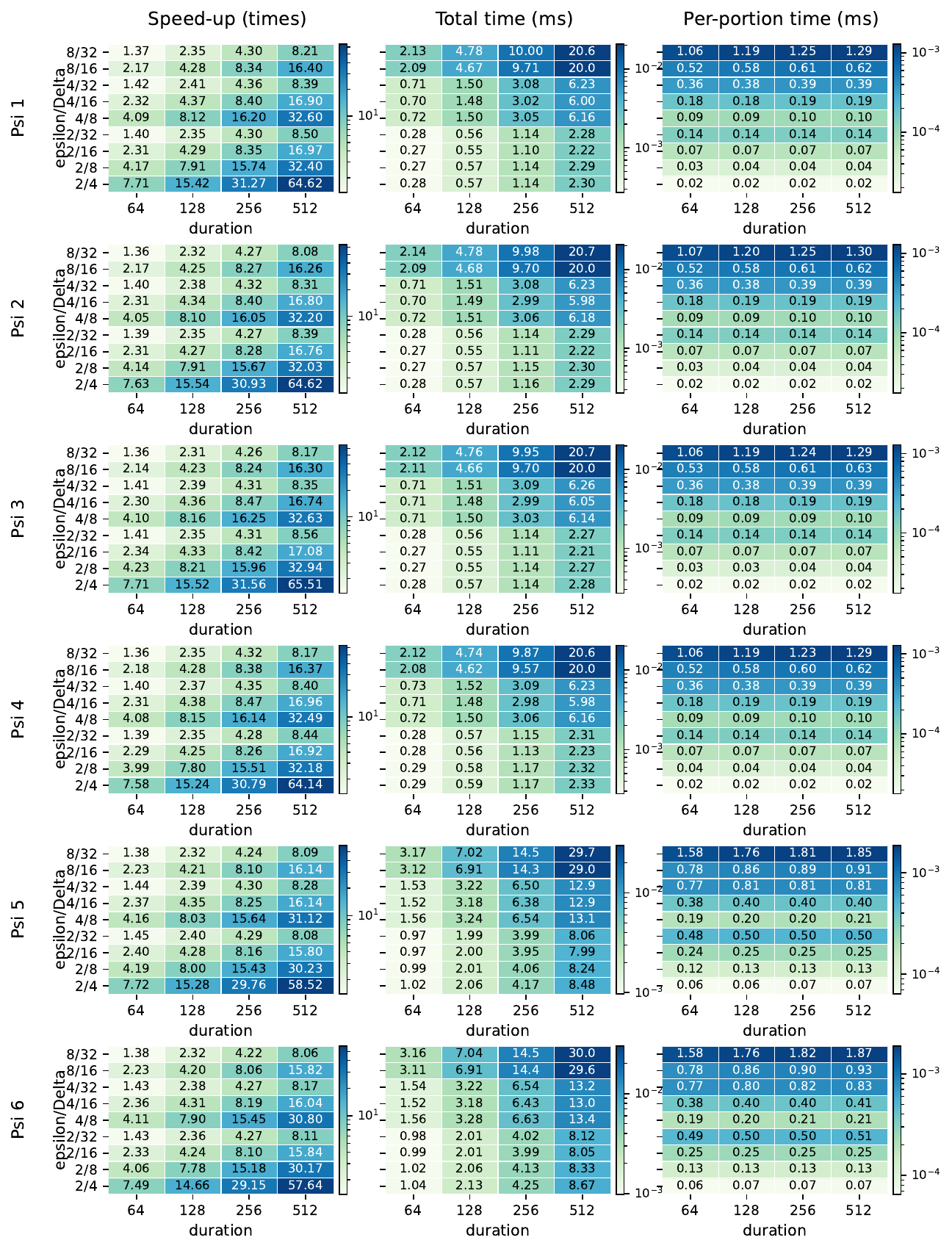}
			\caption{Results on online monitoring $\psi_1$ to $\psi_6$ with \textsc{Adm-Incr} on distributed traces created by the RG. The speedup values are against \textsc{Adm-Naive}.}	\label{fig:rgresultsOnline}
		\end{center}
	\end{figure*}

	\paragraph*{Random Generator, Online Monitoring.}
	Figure~\ref{fig:rgresultsOnline} summarizes the \textsc{Adm-Incr} results for $\psi_1$ to $\psi_6$ on RG traces; the left column shows the speedup over \textsc{Adm-Naive}, the middle column shows the total monitoring time, and the right column shows the corresponding time per portion.
	For each heatmap, columns give the trace duration $d$, and rows give the evaluated $(\varepsilon,\Delta)$ configurations.

	As expected, \textsc{Adm-Incr} is consistently faster than \textsc{Adm-Naive}, with speedups between $1.36\times$ and $65.51\times$.
	If a trace arrives in $N = d/\Delta$ portions, the expected speedup of \textsc{Adm-Incr} against \textsc{Adm-Naive} is about $N/2 = d/(2\Delta)$, as seen in the results.
	For example, when $d = 512$, $\varepsilon = 2$, and $\Delta = 4$, the speedup ranges from $57\times$ to $66\times$ across the six formulas.
	For each formula, increasing $d$ while keeping  $\varepsilon$ and $\Delta$ results in nearly constant mean time per portion for \textsc{Adm-Incr}, while it leads to a roughly linear growth for \textsc{Adm-Naive}.
	Note that the slight increase with trace duration is due to the first portion requiring less work and having less effect on the average for longer traces.

	Portion length has little effect on the total \textsc{Adm-Incr} time.
	For example, when $d = 512$ and $\varepsilon = 2$, its total time varies by less than $8\%$ across the tested values of $\Delta$.
	Nonetheless, portion length determines per-portion time since the monitor can process smaller portions faster.
	Each doubling of $\Delta$ roughly halves the speedup because \textsc{Adm-Naive} recomputes the observed prefix only half as often.
	Clock skew has a stronger effect on the running times, but the effect depends on the formula.
	When $d = 512$ and $\Delta = 16$, doubling $\varepsilon$ multiplies the running times of both \textsc{Adm-Incr} and \textsc{Adm-Naive} by factors between $1.6$ and $3.3$ across all formulas.
	Nonetheless, the results suggest that \textsc{Adm-Incr} can handle high-frequency inputs, since its mean time per portion remains below $2\,\mathrm{ms}$.

\begin{figure*}[t]
			\begin{center}
				\includegraphics[width=\linewidth]{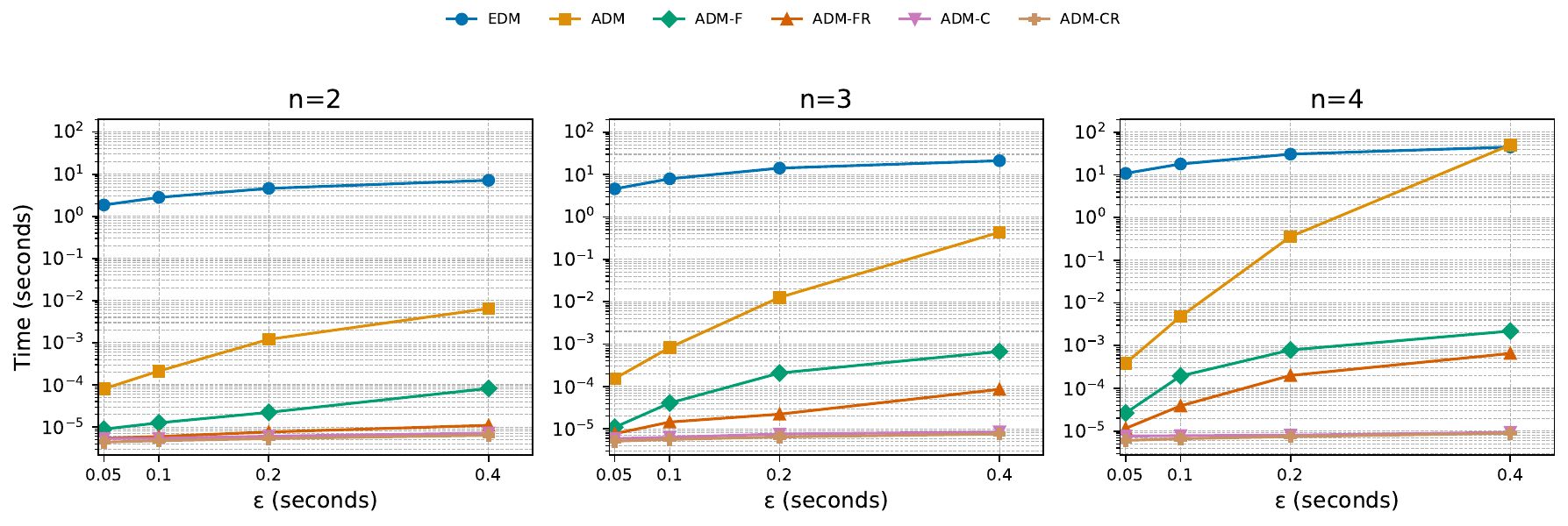}
				\caption{Mean running times for monitoring $\varphi_{\text{WT}}$ on the $1\,\mathrm{s}$, $20\,\mathrm{Hz}$ water-tank trace. Each point is the mean of 10 runs, and the vertical axis is logarithmic.}
				\label{fig:wtresults}
			\end{center}
		\end{figure*}

	\paragraph*{Water Tank.}
	Figure~\ref{fig:wtresults} shows the running times for monitoring $\varphi_{\text{WT}}$.
	No configuration timed out.
	\textsc{Edm} took between $1.86$ and $45.01$ seconds.
	Plain \textsc{Adm} scaled poorly and reached $50.25$ seconds, becoming slower than \textsc{Edm} for $n = 4$ and $\varepsilon = 0.4$.
	The numerical abstractions removed most of this growth:
	\textsc{Adm-F} and \textsc{Adm-Fr} remained below $2.20\,\mathrm{ms}$ and $658\,\mu\mathrm{s}$, respectively, while \textsc{Adm-C} and \textsc{Adm-Cr} remained below $10\,\mu\mathrm{s}$ in every configuration.
	Using a reference clock made \textsc{Adm-Fr} between $1.4\times$ and $9.3\times$ faster than \textsc{Adm-F}, but provided only a modest additional improvement for the already inexpensive coarse abstraction.
	\textsc{Adm-Cr} had the lowest mean running time in every configuration, with speedups from $420{,}000\times$ to $5{,}100{,}000\times$ over \textsc{Edm}.
	These results show that the specialized numerical abstractions avoid the poor scaling of asynchronous products.

\begin{figure*}[t]
			\begin{center}
				\includegraphics[width=\linewidth]{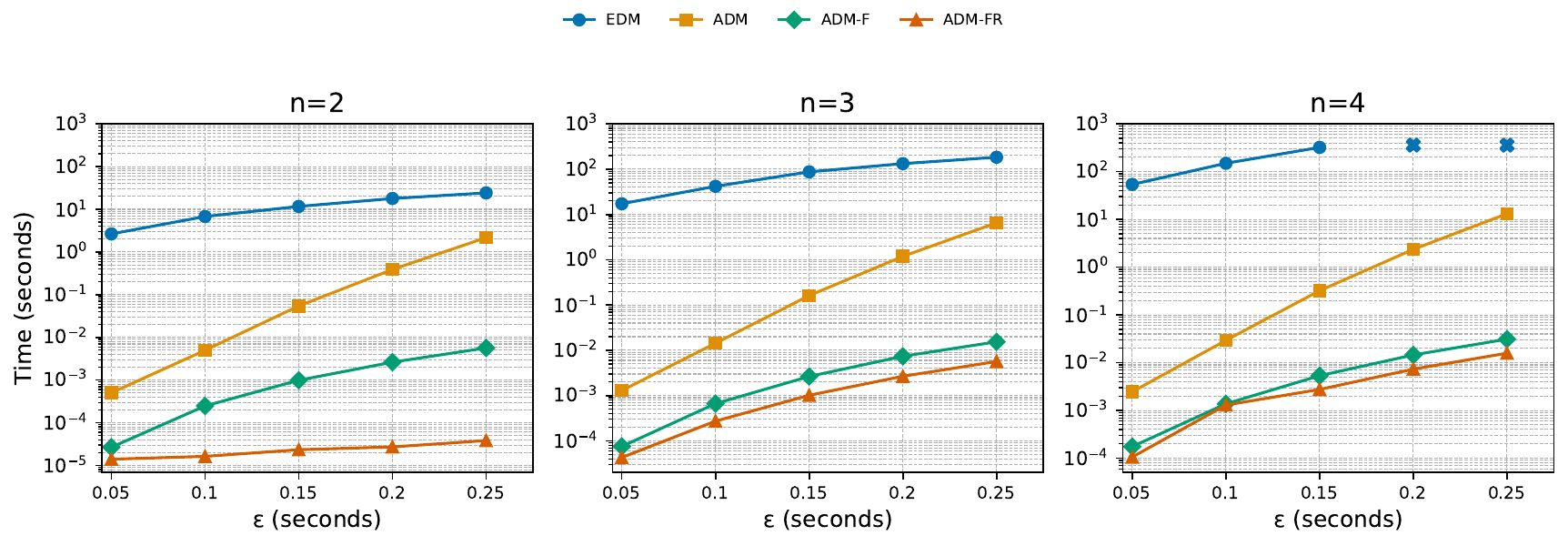}
				\caption{Mean running times for monitoring $\varphi_{\text{SD}}$ on the $1\,\mathrm{s}$, $20\,\mathrm{Hz}$ drone trajectories. Each point is the mean of 10 runs, and the vertical axis is logarithmic. Crosses mark the two configurations where \textsc{Edm} exceeded the $360\,\mathrm{s}$ time limit.}
				\label{fig:sdresults}
			\end{center}
		\end{figure*}

	\paragraph*{Swarm of Drones.}
	Figure~\ref{fig:sdresults} shows the running times for monitoring $\varphi_{\text{SD}}$.
	\textsc{Edm} took between $2.65$ and $322.17$ seconds in the 13 configurations that it completed; it exceeded the time limit for $n = 4$ with $\varepsilon \in \{0.20, 0.25\}$.
	Plain \textsc{Adm} was faster than \textsc{Edm} in every completed configuration, but its running time grew to $13.16$ seconds.
	\textsc{Adm-F} reduced the running time of \textsc{Adm} by between $14\times$ and $425\times$, and \textsc{Adm-Fr} was faster still in every configuration.
	The largest gains for the relative method occurred for two drones; for three or four drones, \textsc{Adm-Fr} was between $1.1\times$ and $2.8\times$ faster than \textsc{Adm-F}.
	Overall, \textsc{Adm-Fr} had the lowest mean running time in every configuration, ranging from $14\,\mu\mathrm{s}$ to $15.8\,\mathrm{ms}$.
	Over the completed \textsc{Edm} configurations, its speedups ranged from $32{,}000\times$ to $658{,}000\times$.
	\textsc{Adm-C} and \textsc{Adm-Cr} are not applicable because squared differences are nonmonotone in the individual coordinates.
		
	\paragraph*{Summary.}
	To answer RQ1, the approximate monitors are much faster than \textsc{Edm} in almost all configurations, with speedups up to six orders of magnitude, but they may return $?$ when \textsc{Edm} gives a conclusive verdict.
	For the untimed formulas, \textsc{Adm}'s extra inconclusive rate is low because the signal values at the start or end determine the verdict.
	For the timed response formulas, however, it returns $?$ on many violations even when \textsc{Edm} is conclusive, because the profile construction can make a response appear possible when no single retiming results in one.

	For RQ2, the combined monitor returns the same verdict as \textsc{Edm} while reducing the estimated total time by $21.4\%$.
	It is faster in every configuration except the two in which \textsc{Adm} is inconclusive on every trace.
	
	For RQ3, \textsc{Adm-Incr} scales well.
	When clock skew and portion length are fixed, its mean time per portion is nearly constant.
	It is faster than \textsc{Adm-Naive} in every configuration, with speedups from $1.36\times$ to $65.51\times$.

	\section{Conclusion} \label{sec:conclusion}
	
	We introduced an approximate and modular framework for distributed monitoring of STL under partial synchrony, covering offline monitoring of future- and past-time STL and online monitoring of past-time STL.

	In the offline setting, the proposed method achieves several orders-of-magnitude speedups over exact SMT-based distributed STL monitors.
	This gain comes at the cost of precision, although conclusive verdicts remain sound.
	Our evaluation shows that the loss of precision depends strongly on the temporal formula and the maximum skew between local clocks, and identifies classes of specifications for which approximate monitors remain precise in practice.
	We also propose combining approximate and exact monitoring in a way that still yields significant efficiency gains while preserving the full precision of the exact monitor.
	In the online setting, we presented an incremental algorithm for past-time STL.
	Our experiments show that, for fixed clock skew and portion length, its average processing time per portion remains low and nearly constant as the trace grows.

	In future work, we will exploit the modular structure of our monitors to achieve finer control over the efficiency-precision tradeoff.
	More specifically, for each operator, we can instantiate either the exact or the approximate evaluation algorithm.

	\clearpage
	\clearpage
	\clearpage
	\clearpage

		\backmatter

	\bmhead{Acknowledgments}
	We thank the anonymous reviewers for their helpful comments.

	\section*{Statements and Declarations}

	\bmhead{Funding}
	This work was supported in part by the European Research Council (grant agreements Nos.~101020093 and 101055412), the U.S. National Science Foundation (award CCF-2118356), and the Chips Joint Undertaking through the A-IQ Ready project (grant agreement No.~101096658).

	\bmhead{Competing interests}
	The authors have no relevant financial or non-financial interests to disclose.

	\bmhead{Data and code availability}
	The implementations of \textsc{Adm}, its evaluated variants, and \textsc{Edm}, together with the distributed traces, case-study inputs, raw timing and verdict data, processed results, experiment and plotting scripts, environment specifications, and reproduction instructions, are available in the repository at \url{https://github.com/egesarac/ApxDistMon}.
	The materials are released under the MIT License.
	
	\bibliographystyle{plain}
	\bibliography{main}
\end{document}